\documentclass[aip,jcp,reprint,eqsecnum,superscriptaddress,twocolumn,longbibliography,floatfix]{revtex4-2}

\usepackage{empheq}
\usepackage{graphicx}
\usepackage{dcolumn}
\usepackage{bm}

\usepackage[utf8]{inputenc}
\usepackage[T1]{fontenc}
\usepackage{mathptmx}
\usepackage{etoolbox}
\usepackage{xcolor}
\usepackage{mathpazo}
\usepackage{bm}
\usepackage[unicode=true,
 bookmarks=true,bookmarksnumbered=false,bookmarksopen=false,
 breaklinks=false,pdfborder={0 0 0},pdfborderstyle={},backref=false,colorlinks=true]
 {hyperref}
\hypersetup{pdfborderstyle={},pdfborderstyle={},pdfborderstyle={},pdfborderstyle={},pdfborderstyle={},pdfborderstyle={},linkcolor=blue,citecolor=blue}
\usepackage{breakurl}

\usepackage{natbib}
\usepackage{amssymb,amsmath}
\usepackage{comment}

\usepackage{float}

\newcommand{\ex}{\text{ex}}
\newcommand{\ccp}{\text{cp}}
\newcommand{\hr}{\text{HR}}
\newcommand{\bp}{\beta p_\|}
\newcommand{\rr}{\mathbf{r}}
\newcommand{\RR}{\mathbf{R}}
\newcommand{\amin}{a^{\min}}
\newcommand{\td}{\theta_{12}}
\newcommand{\xd}{x_{12}}
\newcommand{\rd}{r_{12}}
\newcommand{\Rd}{R_{12}}
\newcommand{\PP}{\mathcal{P}}
\newcommand{\FF}{\mathcal{F}}
\newcommand{\two}{{(2)}}
\newcommand{\ii}{\mathbf{i}}
\newcommand{\jj}{\mathbf{j}}
\newcommand{\cont}{\text{cont}}

\newcommand\beq{\begin{equation}}
	\newcommand\eeq{\end{equation}}
\newcommand\beqa{\begin{eqnarray}}
	\newcommand\eeqa{\end{eqnarray}}
\newcommand{\nn}{\nonumber\\}
\def\bal#1\eal{\begin{align}#1\end{align}}

\newcommand{\intu}{\int}
\newcommand{\intt}{\int}
\newcommand{\fu}{\phi_u}
\newcommand{\fuo}{\phi_{u_1}}
\newcommand{\fup}{\phi_{u_2}}

\makeatletter
\def\@email#1#2{%
 \endgroup
 \patchcmd{\titleblock@produce}
  {\frontmatter@RRAPformat}
  {\frontmatter@RRAPformat{\produce@RRAP{*#1\href{mailto:#2}{#2}}}\frontmatter@RRAPformat}
  {}{}
}%
\makeatother
\begin{document}

\title[]{Exact anisotropic properties of hard spheres in narrow cylindrical confinement}
\author{Ana M. Montero}
\affiliation{Departamento de F\'isica, Universidad de Extremadura, E-06006 Badajoz, Spain}
\author{Andr\'es Santos}%
 \affiliation{
Departamento de F\'isica and Instituto de Computaci\'on Cient\'ifica Avanzada (ICCAEx), Universidad de Extremadura, E-06006 Badajoz, Spain
}%

\date{\today}

\begin{abstract}
We investigate a quasi-one-dimensional (Q1D) system of hard spheres confined within a cylindrical pore so narrow that only nearest-neighbor interactions occur. By mapping this Q1D system onto a one-dimensional polydisperse mixture of nonadditive hard rods, we obtain exact thermodynamic and structural properties, including the radial distribution function, which had remained elusive in previous studies. We derive analytical results for limiting cases, such as small pore diameters, virial expansions, and the high-pressure regime. In particular, we identify a crossover in the anisotropic pressure components: at high densities, the transverse pressure overtakes the longitudinal one when the pore diameter exceeds a critical threshold. We also examine spatial correlations in particle arrangements and radial fluctuations, shedding light on the emergence of ordering in confined systems.
\end{abstract}

\maketitle

\section{\label{sec:introduction}Introduction}
Hard-sphere models offer a simplified yet powerful framework for exploring the fundamental behavior of liquids. They are widely employed in statistical mechanics and molecular simulations to approximate the structural and thermodynamic properties of dense fluids and colloids.\cite{PM86,MU94,WCLSW00,BWQSPP02,RCDRSSV24}

In the study of systems under confined geometries---a field largely driven by advances in nanotechnology---the equilibrium properties of the hard-sphere model have been extensively investigated across a wide range of scenarios, from both theoretical and experimental perspectives. Notable configurations include confinement between two parallel walls,\cite{SL96,HSW97,FD06,MET06,MET07,NSK13,TSAHRD18,NKSCPBV12,SFS14,NSHCPBJK16,JF23,BGM24} spherical confinement,\cite{HYS13,WDW21} and cylindrical confinement in slit pores.\cite{BBW89,ALD96,DSBP98,MP01,KKM08,DG09,HKS09}

Despite their simplicity compared to more complex models, hard-sphere models continue to attract research interest due to their ability to capture key aspects of fluid behavior, including phase transitions\cite{SL96,FD06,MLGORFV14} and transport properties.\cite{GM14,KMS14,WLB20,MGB22,BGM24} Moreover, they serve as a reference system for understanding more intricate interparticle interactions, providing a valuable framework for developing and testing theories of liquid-state physics.

From a theoretical perspective, highly confined systems in slit pores (where the available space along one dimension is much larger than along the other ones) form an interesting class of systems. Similar to purely one-dimensional (1D) systems,\cite{SZK53,K55b,KT68,R71b,P82,BB83a,HC04,S07,BNS09,F16,MS17,MS19,MS20} they can be solved exactly when the interaction is restricted to nearest neighbors.\cite{B62,B64b,KP93} These quasi-one-dimensional (Q1D) systems offer valuable insights into the behavior of confined fluids and represent a significant area of study.\cite{KHD11,M24}

This work focuses on a Q1D system of hard spheres confined in a cylindrical pore, where the narrow pore radius prevents second nearest-neighbor interactions. The exact thermodynamic properties of such systems can be determined using the transfer-matrix method\cite{KMP04,GV13,HFC18} or through approximate approaches.\cite{M18,N19,FS24} However, studying the structural properties beyond purely nearest-neighbor interactions\cite{GV13} remains challenging and is typically addressed through approximations or computer simulations.

To ensure the validity of our exact theoretical framework, we specifically consider the range $0 < \epsilon < \sqrt{3}/2$,  where \(\epsilon\) represents the dimensionless excess pore diameter available to the spheres' centers, expressed in units of the sphere diameter.
In this regime, particles interact exclusively with their first nearest neighbors and can form zigzag configurations near close packing. Wider pores, with $\sqrt{3}/2 < \epsilon < 1$, allow for second-neighbor interactions and the emergence of helical arrangements,\cite{PGO00,C11,MCW11,MCWH12,YB15,FSZSC16,FBSCLC17,MC21,ZSB21,WC23,ZB23} as shown analytically in Ref.~\onlinecite{CWH19}.
Similar close-packed morphologies can also arise in systems with soft interactions.\cite{KT06,OML11}

In this paper, a mapping of the original Q1D system onto a 1D polydisperse mixture of nonadditive hard rods is employed. This approach has previously been applied to a system of Q1D hard disks.\cite{MS23b,MS24b} The theory is extended here to a Q1D hard-sphere fluid, enabling the calculation of both thermodynamic properties (recovering the transfer-matrix results) and structural properties, such as the radial distribution function (RDF), which had remained elusive until now.

This paper is organized as follows. Section~\ref{sec:solution} defines the system under study and its key geometrical properties, along with the mapping used to develop the theoretical solution. Section~\ref{sec:theory} outlines the theoretical framework employed to derive the structural and thermodynamic properties of the system. Section~\ref{sec:limit} applies these methods to obtain analytical results for limiting cases, including very small pore size, very low pressure, and very high pressure. Section~\ref{sec:results} presents the main findings, and Sec.~\ref{sec:conclusions} summarizes the key conclusions.
The most technical steps are presented in  Appendixes \ref{appA}--\ref{sec:discretization}.

\section{\label{sec:solution}The system}

\subsection{\label{sec:systemDefinition}Q1D hard-sphere fluid}

\begin{figure}
	\includegraphics[width=\columnwidth]{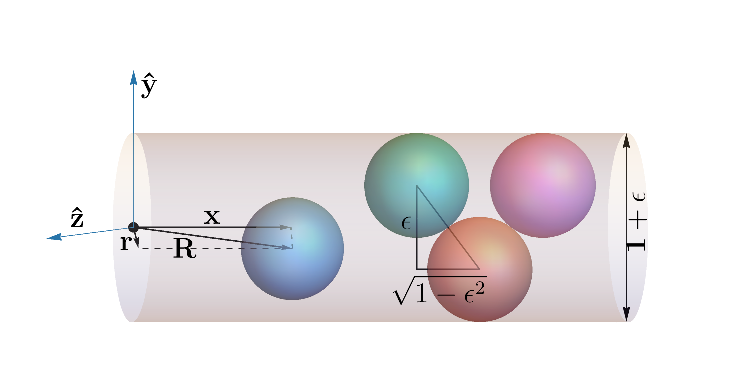}
	\caption{Schematic representation of the confined hard-sphere system. The leftmost particle depicts the system of coordinates and the three rightmost particles represent the close-packing configuration. In this particular example, the value of the excess pore diameter is $\epsilon=0.8$.}
	\label{fig:System}
\end{figure}

Consider a three-dimensional (3D) system of \( N \) hard spheres interacting through the pairwise potential
\begin{equation}
\varphi(R_{12}) = \begin{cases}
	\infty, & R_{12} < 1, \\
	0, & R_{12} > 1,
\end{cases}
\end{equation}
where \( R_{12} = |\RR_{12}| \), with \( \RR_{12} = \RR_1 - \RR_2 \) representing the relative position vector between the centers of two particles. The spheres are assumed to have a unit diameter. The system is confined within a long cylinder of length \( L \gg 1 \) and diameter \( w = 1 + \epsilon \).
To restrict the interactions to nearest neighbors, \(\epsilon\) is limited to the range \( 0 \leq \epsilon \leq {\sqrt{3}}/{2} \simeq 0.866 \).
For simplicity, the cylinder axis is aligned along the \( x \) axis, and the origin of coordinates is defined by any reference point along that axis. Consequently, the position vector of a given sphere is expressed as
\beq
\RR=x\hat{\mathbf{x}}+\rr,\quad \rr=y\hat{\mathbf{y}}+z\hat{\mathbf{z}},
\eeq
with $-\infty<x<\infty$, as shown in Fig.~\ref{fig:System}. In polar coordinates, the two-dimensional vector $\mathbf{r}$ is characterized by its modulus $r$ and the angle $\theta$ so that $y=r\cos\theta$, $z=r\sin\theta$, with  $0\leq r\leq {\epsilon}/{2}$ and $0\leq \theta \leq 2 \pi$.

Given two spheres at positions $\RR_1$ and $\RR_2$, the distance between them is $\Rd=(\xd^2+\rd^2)^{1/2}$, where $\xd=|x_1-x_2|$ is the longitudinal distance and
\beq
\rd=|\rr_1-\rr_2|=\sqrt{r_1^2+r_2^2-2r_1 r_2 \cos\td}
\eeq
is the transverse distance, with $\td = \theta_1 - \theta_2$. When the two spheres are at contact, $\Rd=1$, and then, their longitudinal distance is simply
\beq\label{eq:contact_distance}
a_{\rr_1,\rr_2}=\sqrt{1-\rd^2}.
\eeq

The number density is given by $\rho = N/(L \pi \epsilon^2/4)$,
where only the volume accessible to the particles' centers is considered. Due to the single-file nature of the system, the density can also be characterized by the linear density \(\lambda \equiv N/L\), leading to $\rho = \lambda/(\pi \epsilon^2/4)$.
Since the minimum value of the contact distance in Eq.~\eqref{eq:contact_distance} occurs at \(\theta_{12} = \pi\), the close-packing value of the linear density is $\lambda_{\text{cp}}(\epsilon) = 1/\sqrt{1 - \epsilon^2}$,
as illustrated by the rightmost particles in Fig.~\ref{fig:System}.
Let us denote by
\(P_\|\) and \(P_\perp\)  the longitudinal and transverse pressure components, respectively, so that the mean pressure is given by $P = (P_\| + 2P_\perp)/3$. In what follows, it is convenient to define a 1D analog of the longitudinal pressure as $p_\| =  (\pi \epsilon^2/4)P_\|$.

\begin{figure}
	\includegraphics[width=\columnwidth]{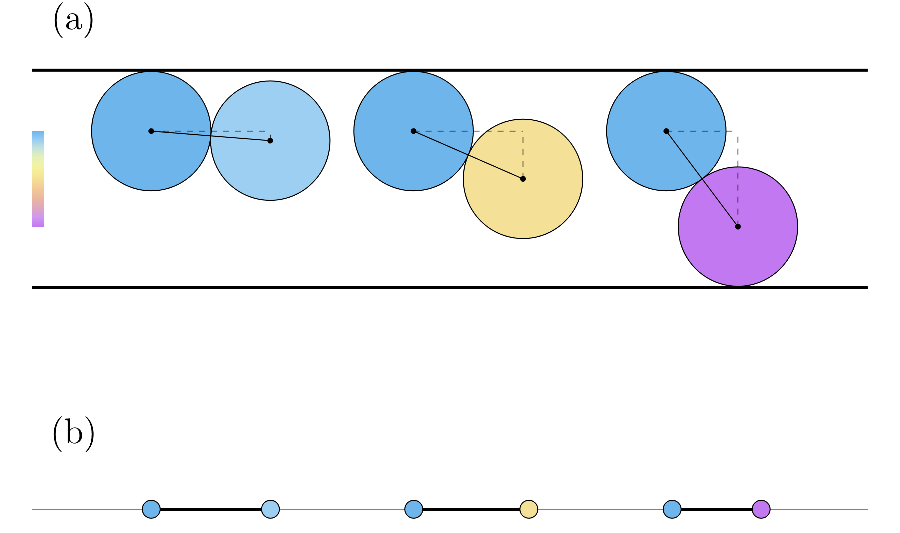}
	\caption{
(a) Q1D system of hard disks confined within a channel that allows a single transverse degree of freedom. Each disk is colored according to its transverse coordinate. Both the transverse and longitudinal components of the contact distance between disks are indicated.
(b) Equivalent 1D mixture obtained by mapping each disk to a  particle on a line. Each circle, colored according to species, represents the center of a 1D particle. The contact distance between a pair of particles (illustrated by a solid line) corresponds to the longitudinal contact distance shown in panel (a).}
	\label{fig:Mapping}
\end{figure}

\subsection{\label{sec:systemMapping}Mapping onto a one-dimensional mixture}

As previously shown for a confined hard-disk system,\cite{MS23b, MS24b} the thermodynamic and structural properties of single-file systems can be determined by mapping the original system onto a polydisperse, nonadditive 1D mixture of hard rods, where all species share the same chemical potential.

As an illustrative example, Fig.~\ref{fig:Mapping} depicts the mapping for a Q1D system of hard disks. In panel (a), the Q1D configuration is shown, with disks colored according to their transverse $y$-coordinate. Although the true distance between two particles at contact is always the same, the longitudinal component of that distance varies depending on the transverse positions of both particles. In the corresponding 1D mixture shown in panel (b), the transverse positional information (i.e., the $y$-coordinate) is encoded in the particle species. As a result, the longitudinal contact distance---the only relevant one in 1D---becomes species-dependent, capturing the geometric constraints of the original Q1D system.

While Fig.~\ref{fig:Mapping} illustrates a system with a single spatially confined dimension, the same rationale can be readily extended to geometries where two spatial dimensions are confined, as in the case of hard spheres inside a cylindrical pore. In such systems, each component of the 1D mixture is characterized by a vector $\rr$, and the hard-core interaction between particles of species $\rr_1$ and $\rr_2$ is specified by a minimum allowed separation $a_{\rr_1, \rr_2}$. The 1D interaction potential is
\beq
\varphi_{\rr_1, \rr_2}(x) =
\begin{cases}
\infty, & x < a_{\rr_1, \rr_2}, \\
0, & x > a_{\rr_1, \rr_2}.
\end{cases}
\eeq
The (negative) nonadditive nature of the mixture is reflected in the fact that \(a_{\rr_1, \rr_2} <  (a_{\rr_1, \rr_1} + a_{\rr_2, \rr_2})/2 = 1\) if $\rr_1\neq \rr_2$.

Note that, within this 1D framework, only \(\lambda\) and \(p_\|\) have physical significance. However, there is a one-to-one correspondence between these quantities and their original 3D counterparts, which allows all properties of the 3D system to be effectively derived from the 1D model.

\section{\label{sec:theory}Exact theoretical solution}

Consider the mapped 1D mixture of hard rods, where \(\phi^2_\rr\) denotes the composition distribution function of the polydisperse mixture. Here, \(\phi^2_\rr d^2\rr\) represents the fraction of particles belonging to a species with a label comprised between \(\rr\) and \(\rr + d\rr\). In the original 3D system, this same quantity corresponds to the probability of finding a particle within an elementary cross section \(d^2\rr\) at a transverse vector \(\rr\).
If the chemical potential of all species in the mixture is the same, the composition distribution function \(\phi^2_\rr\) is not a free parameter but is determined by the solution of the following eigenfunction problem:\cite{MS23b}
\beq
\label{eq:eigenfunction}
\int d^2\rr_2\,e^{-a_{\rr_1,\rr_2}\bp}\phi_{\rr_2}=\ell \phi_{\rr_1},
\eeq
where $\beta = 1/k_B T$ is the inverse temperature, with $k_B$ being the Boltzmann constant, and $\ell$ is the largest eigenvalue, which is related to the excess free energy and chemical potential.\cite{MS23b}

Due to the cylindrical symmetry of the confining channel, \(\phi^2_\rr\) depends only on the radial distance \(r\). Consequently, the normalization condition becomes
\beq
\label{eq:normalization_1}
\int d^2\rr\,\phi^2_\rr=\pi\int_0^{\frac{\epsilon^2}{4}} du\,\phi^2_u=1,
\eeq
where \(u \equiv r^2\) and the notation \(\phi_\rr \to \phi_u\) has been introduced.
Analogously, Eq.~\eqref{eq:eigenfunction} can be  rewritten as
\beq
\label{eq:eigenfunction2}
\frac{1}{2}\int_0^{\frac{\epsilon^2}{4}} du_2\,\fup\int_0^{2\pi}d\td\,e^{-a_{\rr_1,\rr_2}\bp}=\ell\fuo.
\eeq
Note that Eq.~\eqref{eq:eigenfunction2} is equivalent to the one previously obtained via the transfer-matrix method.\cite{KP93,HFC18,V25}
The excess Gibbs--Helmholtz free energy is then obtained as\cite{KP93,MS23b,MS24}
\beq
\label{eq:gex}
\beta g^\ex(\bp,\epsilon)=-\ln\frac{\ell(\bp,\epsilon)}{\pi\epsilon^2/4},
\eeq
where the dependence $\ell = \ell(\bp,\epsilon)$ has been made explicit and we have taken into account that $\ell\to \pi\epsilon^2/4$ in the ideal-gas limit ($\bp\to 0$), as obtained from Eq.~\eqref{eq:eigenfunction2} and the fact that $\fu\to\text{const}$ in that limit.

For the remainder of the text, unless explicitly stated otherwise, the limits of the integrals over the variables $u$ and $\theta$ will be omitted for brevity.

\subsection{\label{sec:thermo}Thermodynamic properties}

Starting from the excess Gibbs--Helmholtz free energy in Eq.~\eqref{eq:gex}, the compressibility factor associated with the longitudinal pressure, $Z_\|\equiv \beta P_\|/\rho=\bp/\lambda$, and the one associated with the transverse one, $Z_\perp\equiv \beta P_\perp/\rho$, can be obtained from their corresponding thermodynamic relations,
\begin{subequations}
\label{eq:zpar&zperp}
\begin{align}\label{eq:zpar}
Z_\|=&1+\bp\left(\frac{\partial \beta g^\ex}{\partial \bp}\right)_\epsilon \nn
=&1+\frac{\pi \bp}{2 \ell}\intu du_1\,\fuo\intu du_2\,\fup\intt d\td\,e^{-a_{\rr_1,\rr_2}\bp}a_{\rr_1,\rr_2},
\end{align}
\begin{align}\label{eq:zperp}
Z_\perp=&1-\epsilon^2\left(\frac{\partial \beta g^\ex}{\partial \epsilon^2}\right)_{\bp} \nn
=&1+\frac{\pi \bp}{4 \ell}\intu du_1\,\fuo\intu du_2\,\fup \intt d\td\,e^{-a_{\rr_1,\rr_2}\bp}\nn
&\times\frac{1-a_{\rr_1,\rr_2}^2}{a_{\rr_1,\rr_2}}.
\end{align}
\end{subequations}
In the derivation of Eq.~\eqref{eq:zperp}, a change of variables
\beq\label{eq:resc}
\overline{u}=\frac{u}{\epsilon^2},\quad \overline{\phi}_{\overline{u}}=\epsilon \fu,\quad \overline{\ell}=\frac{\ell}{\epsilon^2}
\eeq
has been made in order to carry out the evaluation of $Z_\perp=1+(\epsilon/2\overline{\ell})(\partial \overline{\ell}/\partial \epsilon)_{\bp}$.
Moreover, upon  deriving Eq.~\eqref{eq:zpar&zperp}, we have taken into account that
$\intu du_1\,\fuo\intu du_2\,\left(\partial\fup\right) \intt d\td\,\exp(-a_{\rr_1,\rr_2}\bp)=0$,\cite{MS23} where here $\partial$ stands for $\partial_{\bp}$ or $\partial_\epsilon$.
The compressibility factor $Z\equiv \beta P/\rho$ associated with the mean pressure is
\begin{align}
\label{mean_press}
Z=&\frac{1}{3}\left(Z_\|+2Z_\perp\right)\nn
=&1+\frac{\pi \bp}{6 \ell}\intu du_1\,\fuo\intu du_2\,\fup \intt d\td\,\frac{e^{-a_{\rr_1,\rr_2}\bp}}{a_{\rr_1,\rr_2}}.
\end{align}

As proved in Appendix \ref{appA}, Eq.~\eqref{eq:zperp} agrees with the contact value theorem as\cite{FS24}
 \beq
 \label{eq:cont_th}
 Z_\perp=\frac{\pi\epsilon^2}{4}\phi^2_{u=\frac{\epsilon^2}{4}}.
 \eeq

\subsection{Positional fluctuations}

Other relevant quantities are the positional fluctuations of particles relative to the cylindrical pore wall, as characterized by the moments,\cite{V25}
\begin{equation}\label{eq:deltarn}
	\langle(\Delta r)^n\rangle\equiv \pi\intu du\,\left(\frac{\epsilon}{2}-\sqrt{u}\right)^n\phi^2_u.
\end{equation}
In particular, $\langle\Delta r\rangle$ gives the average transverse distance from the wall, i.e., excluding the inaccessible region ${\epsilon}/{2}<r<(1+\epsilon)/{2}$. The standard deviation from this average value is
\beq
\label{sigma_r}
\sigma_{\Delta r}=\sqrt{\langle(\Delta r)^2\rangle-\langle \Delta r\rangle^2}=\sqrt{\langle r^2\rangle-\langle r\rangle^2}.
\eeq
All these quantities provide insight into the spatial distribution of particles within the confined geometry and measure how far particles tend to deviate from the wall of the cylinder, thus playing a crucial role in understanding confinement effects in Q1D systems.

\subsection{\label{sec:struct}Spatial correlations}

Once the composition distribution function $\phi^2_u$ is known at a given $\bp$, the first nearest-neighbor probability distribution function is\cite{S16}
\beq
\label{PP1}
\PP^{(1)}_{\rr_1,\rr_2}(x) = \frac{\bp}{\ell} \frac{\fup}{\fuo} e^{- \bp x} \Theta(x- a_{\rr_1,\rr_2}),
\eeq
where $\Theta(\cdot)$ denotes the Heaviside step function. Due to the cylindrical symmetry, $\PP^{(1)}_{\rr_1,\rr_2}(x)$ depends on the vectors $\rr_1$ and $\rr_2$  only through $u_1$, $u_2$, and the relative angle $\td$.
By using Eq.~\eqref{eq:eigenfunction2}, one can see that the first nearest-neighbor probability distribution is correctly normalized,
\beq
\label{norma1}
\frac{1}{2}\intu  du_2\intt d \theta_{12}  \int_0^\infty dx\,\PP^{(1)}_{\rr_1,\rr_2}(x)=1.
\eeq
Higher order nearest-neighbor distributions are computed by convoluting $\PP^{(1)}_{\rr_1,\rr_2}(x)$:
\beq\label{eq:convP1}
\PP^{(n)}_{\rr_1,\rr_2}(x)=\frac{1}{2}\intu du_3\intt d\theta_{13}\int_0^x dx'\,\PP^{(n-1)}_{\rr_1,\rr_3}(x')
\PP^{(1)}_{\rr_3,\rr_2}(x-x').
\eeq
Note that $\PP^{(n)}_{\rr_1,\rr_2}(x)$ also satisfies the normalization condition [Eq.~\eqref{norma1}]. The simplest example of Eq.~\eqref{eq:convP1} is the second-neighbor probability distribution, which reads
\beq
	\PP^{(2)}_{\rr_1,\rr_2}(x)=
\frac{1}{2}\left(\frac{\bp}{\ell}\right)^2\frac{\phi_{u_2}}{\phi_{u_1}}e^{-\bp x}\FF_{\rr_1,\rr_2}(x),
\eeq
where
\bal
\FF_{\rr_1,\rr_2}(x)=&\intu du_3\intt d\theta_{13}\left(x-a_{\rr_1,\rr_3}-a_{\rr_3,\rr_2}\right)
\nn&\times
\Theta\left(x-a_{\rr_1,\rr_3}-a_{\rr_3,\rr_2}\right).
\eal
is a purely geometric function that vanishes in the region $x\leq a^\two_{\rr_1,\rr_2}\equiv\min_{\rr_3}\{a_{\rr_1,\rr_3}+a_{\rr_3,\rr_2}\}$.
In particular, if $u_1=u_2=u$,  one has $a^\two_{\rr_1,\rr_2}=2a_{\rr_1,\rr_3}$ with $u_3={\epsilon^2}/{4}$ and $\theta_{13}={\theta_{12}}/{2}-\pi$, i.e.,
\beq
\label{atwo}
\left.a^\two_{\rr_1,\rr_2}\right|_{u_1=u_2=u}=2\sqrt{1-u-\frac{\epsilon^2}{4}-\epsilon\sqrt{u}\cos\frac{\theta_{12}}{2}}.
\eeq

In terms of the probability distribution functions $\PP_{\rr_1,\rr_2}^{(n)}(x)$, the component--component RDF in the 1D mixture is given by
\beq
\label{eq:gpartial}
g_{\rr_1,\rr_2}(x) = \frac{1}{\lambda \phi^2_{u_2}} \sum_{n=1}^\infty \PP^{(n)}_{\rr_1,\rr_2}(x),
\eeq
while the total longitudinal RDF is
\beq
\label{eq:gtotal}
g(x)=\frac{\pi}{2}\intu du_1\,\phi^2_{u_1}\intu du_2\,\phi^2_{u_2}\intt d\td\,g_{\rr_1,\rr_2}(x).
\eeq
In the original 3D system, the function $g_{\rr_1,\rr_2}(x)$ is related to  the probability density of finding a pair of particles with transverse positions $\rr_1$ and $\rr_2$ at a longitudinal distance $x$, independently of which neighbor they are.

From Eqs.~\eqref{PP1} and \eqref{eq:gpartial} we can obtain the contact value $g_{\rr_1,\rr_2}^\cont=g_{\rr_1,\rr_2}(a_{\rr_1,\rr_2}^+)$ as
\beq
\label{contact}
g_{\rr_1,\rr_2}^\cont=\frac{Z_\|}{\ell\phi_{u_1}\phi_{u_2}}e^{-a_{\rr_1,\rr_2}\bp}.
\eeq

One can also derive the expression of $g(x)$ at $x=1$ since, at that point, only the first nearest neighbors contribute. Setting $x=1$ in Eq.~\eqref{PP1} we obtain
\beq
g(1)=\frac{Z_\|}{\ell}e^{-\bp}\left(\pi \intu du\,\fu\right)^2.
\eeq

The convolution structure of Eq.~\eqref{eq:convP1} suggests the introduction of the Laplace transforms,
\begin{subequations}
\beq
\label{eq:Omega01}
\Omega_{\rr_1,\rr_2}(s) = \int_0^{\infty}dx\, e^{-s x} e^{- \beta\varphi_{\rr_1,\rr_2}(x)}= \frac{e^{-a_{\rr_1,\rr_2}s}}{s},
\eeq
\beq
\label{eq:PP1s}
\widetilde{\PP}^{(1)}_{\rr_1,\rr_2}(s) = \frac{\bp}{\ell} \frac{\fup}{\fuo} \Omega_{\rr_1,\rr_2}(s + \bp),
\eeq
\begin{align}
\label{3.13c}
\widetilde{\PP}^{(n)}_{\rr_1,\rr_2}(s) =&\frac{1}{2}\intu du_3\intt d\theta_{13}\,\widetilde{\PP}^{(n-1)}_{\rr_1,\rr_3}(s)\widetilde{\PP}^{(1)}_{\rr_3,\rr_2}(s) \nn
=& \left(  \left[\widetilde{\mathsf{P}}^{(1)}(s)\right]^n \right)_{\rr_1,\rr_2}.
\end{align}
\end{subequations}
In the second step of Eq.~\eqref{3.13c}, $\widetilde{\mathsf{P}}^{(1)}(s)$ denotes the matrix with elements $\widetilde{\PP}^{(1)}_{\rr_1,\rr_2}(s)$ and the standard definition for matrix multiplication of infinite-dimensional matrices (analogous to the finite case) has been applied.
Inserting Eq.~\eqref{3.13c} into the Laplace transform of Eq.~\eqref{eq:gpartial}, one gets
\beq
\label{eq:Gs01}
\widetilde{G}_{\rr_1,\rr_2}(s)=
\frac{1}{\lambda\fup^2}\left(\widetilde{\mathsf{P}}^{(1)}(s)\cdot\left[\mathsf{I}-\widetilde{\mathsf{P}}^{(1)}(s)\right]^{-1}\right)_{\rr_1,\rr_2},
\eeq
where the $(\rr_1,\rr_2)$ element of the unit matrix $\mathsf{I}$ is the Dirac delta $\delta(\rr_1-\rr_2)$.
Equation \eqref{eq:Gs01} is not but the formal solution to the integral equation,
\begin{align}
\label{eq:GsLaplace}
\frac{\Omega_{\rr_1,\rr_2}(s+\bp)}{\lambda\fuo}=&\frac{\ell\fup}{\bp} \widetilde{G}_{\rr_1,\rr_2}(s)-\int d^2\rr_3\, \phi_{u_3}\widetilde{G}_{\rr_1,\rr_3}(s)\nn
&\times\Omega_{\rr_3,\rr_2}(s+\bp).
\end{align}
The Laplace transform of the total pair correlation function is then
\beq
\label{eq:gtotalLaplace}
\widetilde{G}(s)=\frac{\pi}{2}\intu du_1\,\phi^2_{u_1}\intu du_2\,\phi^2_{u_2}\int d\td\,\widetilde{G}_{\rr_1,\rr_2}(s).
\eeq

Going back to the original 3D confined system, defining a global RDF, $g(R)$, is not as straightforward as it was for its longitudinal counterpart in Eq.~\eqref{eq:gtotal}, due to the loss of translational invariance---which is preserved only along the $x$-direction. However, it is still possible to define a nominal RDF, denoted $\hat{g}(R)$, such that $2\lambda \hat{g}(R) dR$ represents the average number of particles at a distance between $R$ and $R + dR$ from a reference particle. If we define the local number density as $n_1(\RR) = \lambda \phi_u^2$, the function $\hat{g}(R)$ can be obtained from the two-body configurational distribution $n_2(\RR_1, \RR_2) = n_1(\RR_1) n_1(\RR_2) g_{\rr_1, \rr_2}(\xd)$ as
\begin{align}\label{eq:nhat0}
\hat{g}(R)=&\frac{N^{-1}}{2\lambda}\int d^3\RR_1 \int d^3\RR_2 \,n_2(\RR_1,\RR_2)\delta\left(R-R_{12}\right)\nn
=&\frac{\pi}{2}\int_0^L d\xd \intu du_1\,\fuo^2 \intu du_2\,\fup^2 \nn
&\times\intt d\td\, g_{\rr_1,\rr_2}(\xd) \delta\left(R-\sqrt{\rd^2+\xd^2}\right).
\end{align}

Using the identity
\beq
\delta\left(R-\sqrt{\rd^2+\xd^2}\right)=\frac{R}{\xd}\delta\left(\xd-\sqrt{R^2-\rd^2}\right),
\eeq
Eq.~\eqref{eq:nhat0} transforms into
\beq
\label{eq:nhat}
\hat{g}(R)=\frac{\pi}{2} \intu du_1\,\fuo^2 \intu du_2\,\fup^2 \intt d\td\,\hat{g}_{\rr_1,\rr_2}(R),
\eeq
where
\beq
\hat{g}_{\rr_1,\rr_2}(R)\equiv \frac{R}{\sqrt{R^2-\rd^2}}g_{\rr_1,\rr_2}\left(\sqrt{R^2-\rd^2}\right).
\eeq
In Eq.~\eqref{eq:nhat}, it is understood that $R\geq 1>\epsilon\geq \rd$ since $\hat{g}(R)=0$ if $R<1$.
Note that, if $R\gg \rd$, we can expand $\hat{g}_{\rr_1,\rr_2}(R)$ in powers of $\rd$,
\bal
\hat{g}_{\rr_1,\rr_2}\left(R\right)=&g_{\rr_1,\rr_2}\left(R\right)+q^{(1)}_{\rr_1,\rr_2}\left(R\right)\frac{r_{12}^2}{2R^2}
+q^\two_{\rr_1,\rr_2}\left(R\right)\frac{r_{12}^4}{8R^4}
+\cdots,
\eal
where
\begin{subequations}
\beq
q^{(1)}_{\rr_1,\rr_2}\left(R\right)\equiv 2g_{\rr_1,\rr_2}\left(R\right)-\partial_R\left[R g_{\rr_1,\rr_2}\left(R\right)\right],
\eeq
\bal
q^\two_{\rr_1,\rr_2}\left(R\right)\equiv&
8g_{\rr_1,\rr_2}\left(R\right)-7\partial_R\left[R g_{\rr_1,\rr_2}\left(R\right)\right] \nn
 &+\partial_R^2\left[R^2 g_{\rr_1,\rr_2}\left(R\right)\right].
\eal
\end{subequations}

\section{\label{sec:limit}Limiting behaviors}

When studying a complex system---especially one lacking a fully analytical solution---analytical results in limiting-case scenarios serve as reliable reference points for validating numerical or approximate methods. They also enhance our understanding of the system by revealing key behaviors. In what follows, we examine several important limiting cases and derive their corresponding asymptotic analytical expressions.

\subsection{\label{sec:smalleps}Limit of small excess pore diameter at fixed $\lambda<1$}

The value of the excess pore diameter $\epsilon$ measures the deviation of the confined 3D system from its pure 1D version at $\epsilon=0$ (in which the Tonks gas behavior is recovered). It is then interesting to analyze how the 3D confined system deviates from the expected Tonks gas as the pore size increases.
Note that the condition $\lambda\leq 1/\sqrt{1-\epsilon^2}$ implies $\epsilon\geq \sqrt{1-\lambda^{-2}}$; thus, the limit $\epsilon\to 0$ is accessible only if $\lambda<1$.

Following the mathematical steps outlined in Appendix~\ref{app:smallepsapp}, one obtains
\begin{subequations}
	\label{eq:phiell_LP}
	\beq\label{eq:philowbp}
	\fu=\frac{2}{\epsilon\sqrt{\pi}}\left[1+\frac{\bp}{2}\left(u-\frac{\epsilon^2}{8}\right)+\cdots\right],
	\eeq
	\beq
	\ell =\frac{\pi\epsilon^2}{4}e^{-\bp}\left(1+\epsilon^2\frac{\bp}{8}+\cdots\right).
	\eeq
\end{subequations}
Inserting these expressions into Eq.~\eqref{eq:zpar&zperp} yields
\begin{subequations}\label{eq:z_smalle}
	\beq
\label{eq:zpar_smalle}
	Z_\|=1+\bp\left(1-\frac{\epsilon^2}{8}+\cdots\right),
	\eeq
	\beq
	Z_\perp=1+\bp\frac{\epsilon^2}{8}+\cdots.
	\eeq
\end{subequations}
Note that, using the Tonks gas equation of state, $\bp = \lambda/(1-\lambda)$, Eq.~\eqref{eq:z_smalle} is consistent with results previously obtained through perturbative methods.\cite{FS24}

The limiting behavior of the longitudinal RDF $\widetilde{G}_{\rr_1, \rr_2}(s)$ can also be studied in the limit $\epsilon \to 0$. As shown in  Appendix~\ref{app:smallepsapp},
\begin{align}\label{eq:Gs_SE}
\widetilde{G}_{\rr_1, \rr_2}(s)=&\widetilde{G}^\hr(s)\bigg[1+\epsilon^2\frac{\lambda}{8} s \widetilde{G}^\hr(s)+\frac{s}{2}\left(1-a_{\rr_1, \rr_2}^2\right)\nn
&-\left(1-e^{-s}\right)\bp \sqrt{u_1 u_2}\cos\td+\cdots\bigg],
\end{align}
where
\beq
\label{eq:GsHR}
\widetilde{G}^\hr(s)=\frac{Z_\| e^{-s}}{s+\bp(1-e^{-s})}
\eeq
is the RDF of pure hard rods in the Laplace space.

While formally correct, Eq.~\eqref{eq:Gs_SE} presents two drawbacks when used to obtain the associated RDF $g_{\rr_1,\rr_2}(x)$. First, it yields $g_{\rr_1,\rr_2}(x) = 0$ in the interval $a_{\rr_1,\rr_2} < x < 1$, where a nonzero value is expected. Second, Eq.~\eqref{eq:Gs_SE} contains a term proportional to $e^{-s}$, whose inverse Laplace transform includes a spurious Dirac delta contribution $\delta(x - 1)$. These two issues are related and can be resolved by rewriting the RDF as
\beq
g_{\rr_1,\rr_2}(x) =
\begin{cases}
  \frac{Z_\|}{\ell\phi_{u_1}\phi_{u_2}}e^{-\bp x}, & a_{\rr_1,\rr_2} < x < 1, \\
  g_{\rr_1,\rr_2}^+(x), & x > 1,
\end{cases}
\label{g12new}
\eeq
where, for small $\epsilon$, the Laplace transform of $g_{\rr_1,\rr_2}^+(x)$ is
\beq
\label{eq:Gs_SEnew}
\widetilde{G}_{\rr_1, \rr_2}^+(s)=\widetilde{G}_{\rr_1, \rr_2}(s)-\frac{Z_\|}{2}\left(1-a_{\rr_1, \rr_2}^2\right)e^{-s}.
\eeq

Using now Eq.~\eqref{eq:gtotal}, we obtain
\beq
g(x) =
\begin{cases}
  g^-(x), & \sqrt{1-\epsilon^2} < x < 1, \\
  g^+(x), & x > 1,
\end{cases}
\label{gnew}
\eeq
where the Laplace transform of $g^+(x)$ is given by
\beq
\widetilde{G}^+(s) = \widetilde{G}^\hr(s) \left\{ 1 + \frac{\epsilon^2}{8} s \left[ 1 + \lambda \widetilde{G}^\hr(s) \right] \right\} - \frac{Z_\|}{8} \epsilon^2 e^{-s}.
\label{Gnew}
\eeq
In particular, for $1 \leq x \leq 3$,
\bal
g^+(x) =& Z_\| \left(1 - \frac{\bp \epsilon^2}{8} \right) e^{-\bp(x - 1)} + Z_\| \bp e^{-\bp(x - 2)}\nn
&\times \left[ \left(1 - \frac{\bp \epsilon^2}{4} \right)(x - 2) + \frac{\epsilon^2}{4} \right] \Theta(x - 2).
\eal
The specific form of $g^-(x)$ is of little relevance, since its domain $\sqrt{1 - \epsilon^2} < x < 1$ has a width of order $\epsilon^2$. Apart from the continuity conditions $g^-(\sqrt{1 - \epsilon^2}) = 0$ and $g^-(1) = g^+(1) = Z_\| (1 - {\bp\epsilon^2}/{8}  + \cdots)$, the Laplace transform of $g^-(x)$ must equal $(Z_\| / 8) \epsilon^2 \exp(-s) + \cdots$ in order to cancel the last term on the right-hand side of Eq.~\eqref{Gnew}. A constructive form is
\beq
g^-(x) = Z_\| \eta^2 \left[ \left(8 - c \epsilon^2 \right) \eta + \frac{c - 6 - \bp}{2} \epsilon^2 \right],
\eeq
where $c$ is a free parameter and $\eta \equiv (x - \sqrt{1 - \epsilon^2}) / \epsilon^2$.

\subsection{\label{sec:smallbp}Limit of small pressure at fixed $\epsilon$}

The limiting behavior at small pressure (or, equivalently, small density) of any given fluid is usually described by the virial expansion. Knowledge of the lowest-order virial coefficients is crucial to understand the behavior of the system.
Although standard virial expansions are typically performed in powers of the density, the free energy and compressibility factor can also be expanded in powers of $\bp$ as
\begin{subequations}
\label{eq:virial}
\beq
\beta g^\ex=\sum_{n=2}^\infty \frac{B_{n\|}'}{n-1}(\bp)^{n-1},
\eeq
\beq
\label{eq:virial1}
Z_\alpha=1+\sum_{n=2}^\infty B_{n\alpha}'(\bp)^{n-1},\quad \alpha= \|\text{ or } \perp.
\eeq
\end{subequations}
Note that the thermodynamic relation in the first equality of Eq.~\eqref{eq:zperp} implies $B_{n\perp}'=-(n-1)^{-1}\epsilon^2\partial B_{n\|}'/\partial \epsilon^2$.
The virial coefficients $B_{n\alpha}$ in the expansions  in powers of $\lambda$ are related to $B_{n\alpha}'$ in a simple way. For instance, $B_{2\alpha}=B_{2\alpha}'$ and $B_{3\alpha}=B_{3\alpha}'+B_{2\alpha}B_{2\|}$.
However, the  truncated expansions in powers of $\bp$ have been shown to perform better for Q1D systems than their counterparts in powers of $\lambda$ and will, therefore, be used here.\cite{MSRH11,M18,MS23}

In the low-pressure regime, we can  write
\begin{subequations}
\label{eq:getB2_0}
	\beq
\label{eq:getB2_0a}
	\fu=\frac{2}{\sqrt{\pi}\epsilon}\left(1+\bp \psi^{(1)}_u+\cdots\right),
	\eeq
	\beq
	\ell=\frac{\pi\epsilon^2}{4}\left(1-\bp B_{2\|}+\cdots\right),
	\eeq
\end{subequations}
where the ideal-gas values (at $\bp=0$) have been determined from Eqs.~\eqref{eq:normalization_1} and \eqref{eq:eigenfunction2}.
Following the mathematical steps outlined in Appendix~\ref{app:smallbpapp}, one obtains
\begin{subequations}
\beq\label{eq:getB2_3}
	\psi^{(1)}_u=-\Psi^\|_u+B_{2\|},
\eeq
\begin{align}\label{eq:B2}
B_{2\|}=\frac{4}{\epsilon^2}\intu du \,\Psi^\|_{u},
\end{align}
\beq\label{eq:B3}
B_{3\|}'=B_{2\|}^2-1+\frac{\epsilon^2}{4}+\frac{8}{\epsilon^2}\intt du\,\psi^{(1)}_u\Psi^\|_{u},
\eeq
\begin{align}\label{eq:B2perp}
B_{2\perp}=\frac{4}{\epsilon^2}\intu du \,\Psi^\perp_{u},
\end{align}
\beq\label{eq:B3perp}
B_{3\perp}'=B_{2\|}B_{2\perp}-\frac{\epsilon^2}{8}+\frac{8}{\epsilon^2}\intt du\,\psi^{(1)}_u\Psi^\perp_{u},
\eeq
\end{subequations}
where the  functions  $\Psi^\|_u$ and $\Psi_u^\perp$ are defined in Eqs.~\eqref{eq:getB2_2} and \eqref{eq:get_B2p1}, respectively.

While the second and third virial coefficients are expressed in terms of  integrals that, to our knowledge, must be performed numerically, explicit expressions can be obtained by expanding in powers of $\epsilon$. The results are
\begin{subequations}
\beq
\label{B2perpExp}
B_{2\|}=1- \frac{\epsilon^2}{2^3} - \frac{5 \epsilon^4}{3\times 2^7}-\frac{7 \epsilon^6}{2^{11}} - \frac{21 \epsilon^8}{2^{14}}-\frac{77 \epsilon^{10}}{2^{17}}
+\mathcal{O}(\epsilon^{12}),
\eeq
\beq
\label{B3perpExp}
B_{3\|}'=-\frac{5 \epsilon^4}{3\times 2^7} - \frac{7 \epsilon^6}{3\times 2^9} - \frac{97 \epsilon^8}{3\times 2^{14}} - \frac{1933 \epsilon^{10}}{15\times 2^{17}} +\mathcal{O}(\epsilon^{12}).
\eeq
\end{subequations}
The expansions of $B_{2\perp}$ and $B_{3\perp}'$ are easily obtained from the relation $B_{n\perp}'=-(n-1)^{-1}\epsilon^2\partial B_{n\|}'/\partial \epsilon^2$.

Equation~\eqref{B2perpExp} coincides with the result derived in Ref.~\onlinecite{M18}. However, the expansion of $B_{3\|}'$ given in Ref.~\onlinecite{M18} differs from the exact result presented in Eq.~\eqref{B3perpExp} already at the leading order (where the exact coefficient $\frac{5}{3}\times 2^{-7}$ is replaced by $2^{-7}$). The origin of this discrepancy lies in the use of standard irreducible diagrams in Ref.~\onlinecite{M18}, which implicitly assumes a cancellation of the so-called reducible diagrams---a cancellation that is not supported in confined systems. A similar problem was already reported in the case of Q1D hard disks.\cite{MS23}

Before closing this subsection, note that, in the limit $\bp\to 0$, the moments and standard deviation defined by Eqs.~\eqref{eq:deltarn} and \eqref{sigma_r} become
\begin{subequations}
\begin{equation}\label{deltarnlowbp}
	\lim_{\bp\to 0}\langle(\Delta r)^n\rangle =\frac{2}{(n+1)(n+2)}\left(\frac{\epsilon}{2}\right)^n.
\end{equation}
\beq
\lim_{\bp\to 0}\sigma_{\Delta r}=\frac{\epsilon}{6\sqrt{2}}.
\eeq
\end{subequations}

\subsection{\label{sec:highbp}Limit of high pressure at fixed $\epsilon$}

In the asymptotic limit $\bp \to \infty$, particles tend to organize into a close-packed arrangement, occupying positions that minimize the distance between the first nearest neighbors. As a result, the minimum value of $a_{\rr_1,\rr_2}$---which directly influences the factor $\exp(-a_{\rr_1,\rr_2}\bp)$ in Eq.~\eqref{eq:eigenfunction2}---becomes critically important. In this high-pressure regime, for given $u_1$ and $u_2$, the function $\exp(-a_{\rr_1,\rr_2}\bp)$ exhibits a sharp maximum at $\td = \pi$. If only $u_1$ is fixed, the maximum occurs at $u_2 = \epsilon^2/4$ and $\td = \pi$. The global maximum of $\exp(-a_{\rr_1,\rr_2}\bp)$ is, therefore, $\exp(-\sqrt{1 - \epsilon^2}\bp)$, attained when $u_1 = u_2 = \epsilon^2/4$ and $\td = \pi$.

As a consequence of the preceding reasoning, one finds that, in the high-pressure regime, the eigenfunction $\fu$ and its eigenvalue $\ell$ adopt the form (see Appendix~\ref{app:highbpapp} for details) as
\begin{subequations}
\label{eq:phiell_HP}
\beq
\label{eq:phi_HP}
\fu \approx\frac{1}{\sqrt{\mathcal{N}_0}}e^{-\amin_u\bp},
\eeq
\begin{align}\label{eq:ell_HP}
\ell\approx \sqrt{\frac{\pi}{2}}\frac{(1-\epsilon^2)^{3/4}}{\epsilon (\bp)^{3/2}}e^{-\sqrt{1-\epsilon^2}\bp},
\end{align}
\end{subequations}
where
\begin{subequations}
\beq
\amin_u=\sqrt{1-\left(\sqrt{u}+\frac{\epsilon}{2}\right)^2},
\eeq
\begin{align}\label{eq:norm_HP}
\mathcal{N}_0\approx\pi  e^{-2\sqrt{1-\epsilon^2}\bp}\frac{\sqrt{1-\epsilon^2}}{2\bp}.
\end{align}
\end{subequations}
This analytical form for the high-pressure limit is analogous to the one in the hard-disk case, in which particles are also arranged in a similar zigzag ordering.\cite{MS23}
From Eq.~\eqref{eq:phi_HP}, one has
\beq
\label{eq:contact}
\phi^2_{u=\frac{\epsilon^2}{4}}\approx\frac{2\bp}{\pi\sqrt{1-\epsilon^2}}.
\eeq

The high-pressure compressibility factors become
\begin{subequations}
	\label{eq:z_HP}
	\beq
	\label{eq:zpar_HP}
	Z_\|\approx \sqrt{1-\epsilon^2}\bp+\frac{5}{2},
	\eeq
	\beq
	\label{eq:zper_HP}
	Z_\perp\approx\frac{\epsilon^2}{2\sqrt{1-\epsilon^2}}\bp-\frac{1}{2}-\frac{3}{4}\frac{\epsilon^2}{1-\epsilon^2}.
	\eeq
\end{subequations}
The subdominant term in Eq.~\eqref{eq:zpar_HP} needs to be retained  if we want to express the limit in terms of the linear density $\lambda$. In that case, Eq.~\eqref{eq:z_HP} can be rewritten as
\begin{subequations}\label{eq:z_HP2}
	\beq\label{eq:z_HP2a}
	Z_\|\approx\frac{\frac{5}{2}}{1-\lambda/\lambda_{\ccp}},
	\eeq
	\beq
	Z_\perp\approx\frac{\frac{5}{4}(\lambda_{\ccp}^2-1)}{1-\lambda/\lambda_{\ccp}}.
	\eeq
\end{subequations}
The factor ${5}/{2}$ in Eq.~\eqref{eq:z_HP2a} was previously observed in Ref.~\onlinecite{V25}.
Since $Z_\perp/Z_\|\to (\lambda_{\ccp}^2-1)/2$ in the limit $\lambda\to\lambda_{\ccp}$, one finds that $Z_\perp>Z_\|$ in that limit only if $\lambda_{\ccp}^2>3$, that is, $\epsilon>\sqrt{{2}/{3}}\simeq 0.816$. This means that $Z_\| > Z_\perp $ for the entire range of densities  if $\epsilon<\sqrt{{2}/{3}}$, whereas for larger pore widths, $Z_\| > Z_\perp $ only up to a certain density, in which case a crossover between both components occurs.

As shown in Appendix~\ref{app:highbpapp}, the high-pressure limits of the positional fluctuation moments and standard deviation are
\begin{subequations}
\beq
\label{eq:deltarn_HP}
\langle(\Delta r)^n\rangle\approx n!\left(\frac{\sqrt{1-\epsilon^2}}{2\epsilon\bp}\right)^{n}
\approx n!\left(\frac{1-\lambda/\lambda_\ccp}{5\epsilon\lambda_\ccp^2}\right)^n,
\eeq
\beq
\label{sigma_r_HP}
\sigma_{\Delta r}\approx \frac{\sqrt{1-\epsilon^2}}{2\epsilon\bp}
\approx \frac{1-\lambda/\lambda_\ccp}{5\epsilon\lambda_\ccp^2}.
\eeq
\end{subequations}
In particular, the second-order moment, $\langle(\Delta r)^2\rangle$, decays as $(\bp)^{-2}$, in agreement with previous numerical evidence.\cite{V25}
It is also notable that $\sigma_{\Delta r}/\langle \Delta r\rangle\to 1$ in the high-pressure limit.

\section{\label{sec:results}Results}

All the results presented in Sec.~\ref{sec:theory}, where the mapped mixture is treated as a 1D mixture with a continuous distribution, are theoretically exact. However, for practical numerical computations, discretization of the system is necessary.\cite{KP93} This involves approximating the polydisperse mixture with a finite, but large, number of discrete components. Consequently, all integrals over the variables $u$ and $\td$ in Sec.~\ref{sec:theory} are replaced by discrete summations. Further details on the numerical procedure can be found in Appendix~\ref{sec:discretization}.
An open-source C++ code used to obtain the results of this section can be accessed from Ref.~\onlinecite{SingleFileHardSpheres}.

\subsection{Compressibility factor}

\begin{figure}
	\includegraphics[width=0.95\columnwidth]{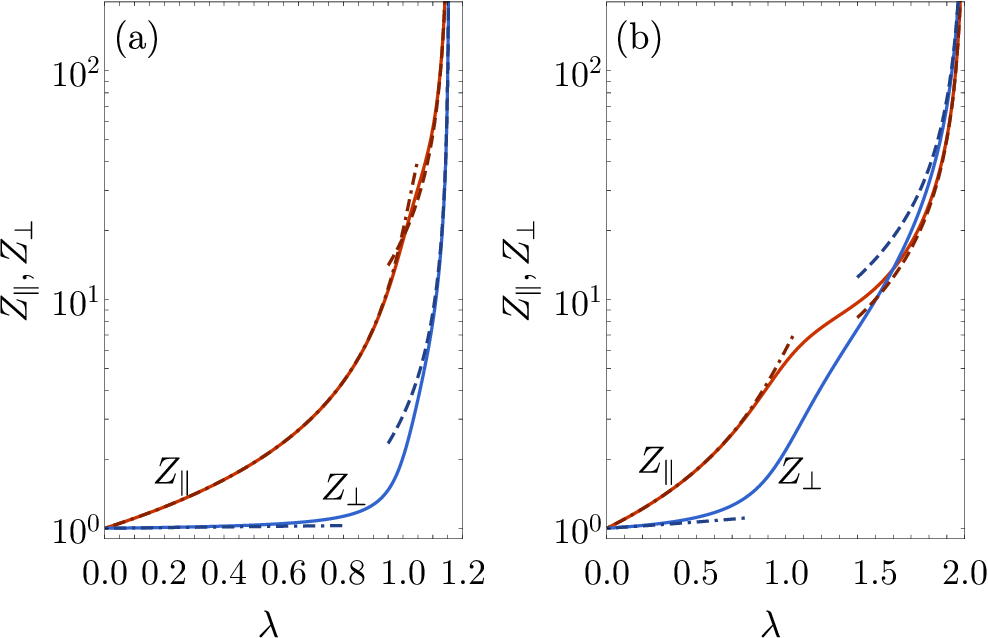}
	\caption{Plot of $Z_\|$ and $Z_\perp$ as functions of the linear density for (a) $\epsilon=0.5$ and (b) $\epsilon=\sqrt{3}/2$. Dashed-dotted lines represent the expansions given by Eq.~\eqref{eq:virial1} truncated after the third virial coefficient, while dashed lines represent the high-pressure behavior given by Eq.~\eqref{eq:z_HP2}.}
	\label{fig:Eos}
\end{figure}

Because of the pronounced anisotropy of the system, the longitudinal ($Z_\|$) and transverse ($Z_\perp$) components of the compressibility factor must be studied separately. Figure~\ref{fig:Eos} shows these quantities, along with their corresponding low- and high-pressure approximations.

The virial expansions given by Eq.~\eqref{eq:virial1} remain highly accurate up to medium-range densities, even when truncated after the third virial coefficient. For both a pore size of $\epsilon=0.5$ and the maximum pore size, $\epsilon = \sqrt{3}/2$, the approximation yields values of $Z_\|$ that are essentially indistinguishable from the exact solution up to $\lambda \simeq 1.0$, which corresponds to $\lambda/\lambda_\ccp \simeq 0.87$ and $\lambda/\lambda_\ccp \simeq 0.5$ for $\epsilon=0.5$ and $\sqrt{3}/2$, respectively.

The high-pressure approximations in Eq.~\eqref{eq:z_HP2} also provide very good results over a reasonable range of large densities, especially for lower values of $\epsilon$.

\begin{figure}
	\includegraphics[width=\columnwidth]{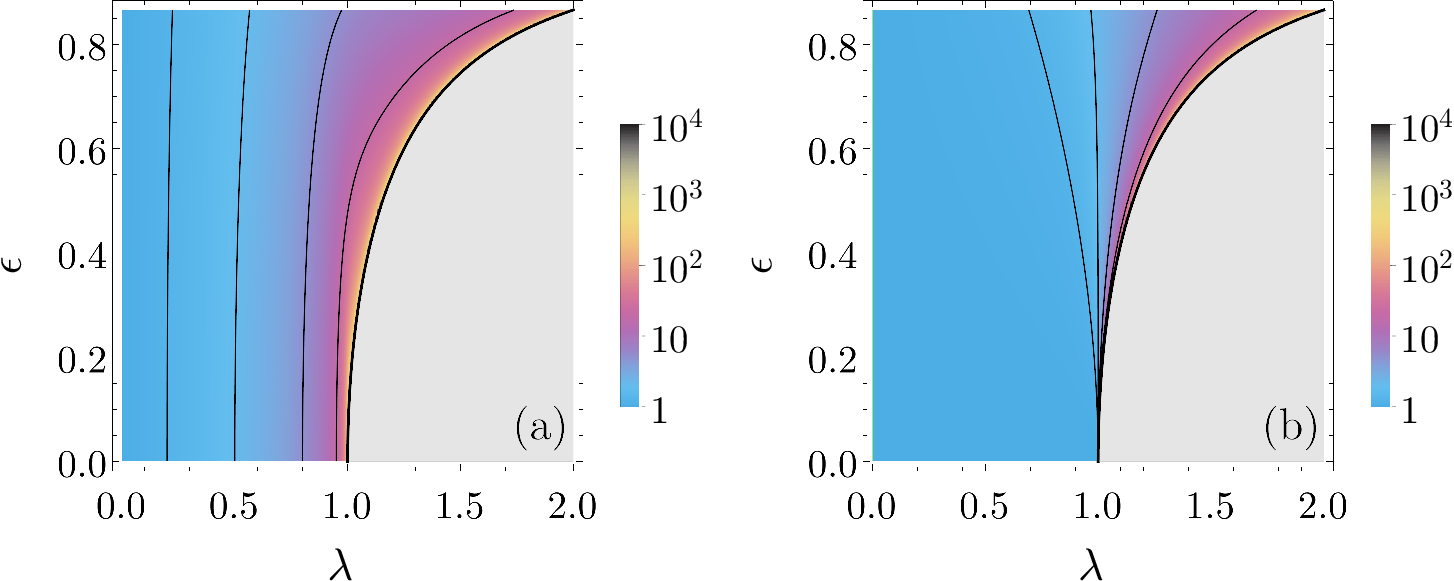}
	\caption{Contour plots of (a) $Z_\|$ and (b) $Z_\perp$ as functions of $\lambda$ and $\epsilon$. In each panel, the contour lines correspond, from left to right, to the values $Z_{\|,\perp} = 1.25$, $2$, $5$, and $20$.}
	\label{fig:EosMap}
\end{figure}

It is interesting to note that, as expected from the results in Sec.~\ref{sec:highbp}, no crossover between $Z_{\|}$ and $Z_\perp$ occurs when $\epsilon=0.5<\sqrt{2/3}$. In contrast, for $\epsilon=\sqrt{3}/2$, $Z_\perp < Z_\|$ only up to a certain density ($\lambda\simeq 1.6$), where both components cross.

Figure~\ref{fig:Eos} is complemented by Fig.~\ref{fig:EosMap}. We observe that, for a given value of $\lambda$, the longitudinal compressibility factor $Z_\|$ decreases as $\epsilon$ increases, with this effect becoming more pronounced at higher densities. In the case of the transverse compressibility factor $Z_\perp$, a qualitatively similar trend is seen for linear densities larger than $\lambda \approx 1$. However, for smaller values of $\lambda$, $Z_\perp$ increases with increasing $\epsilon$.

\subsection{Positional fluctuations}
\begin{figure}
	\includegraphics[width=\columnwidth]{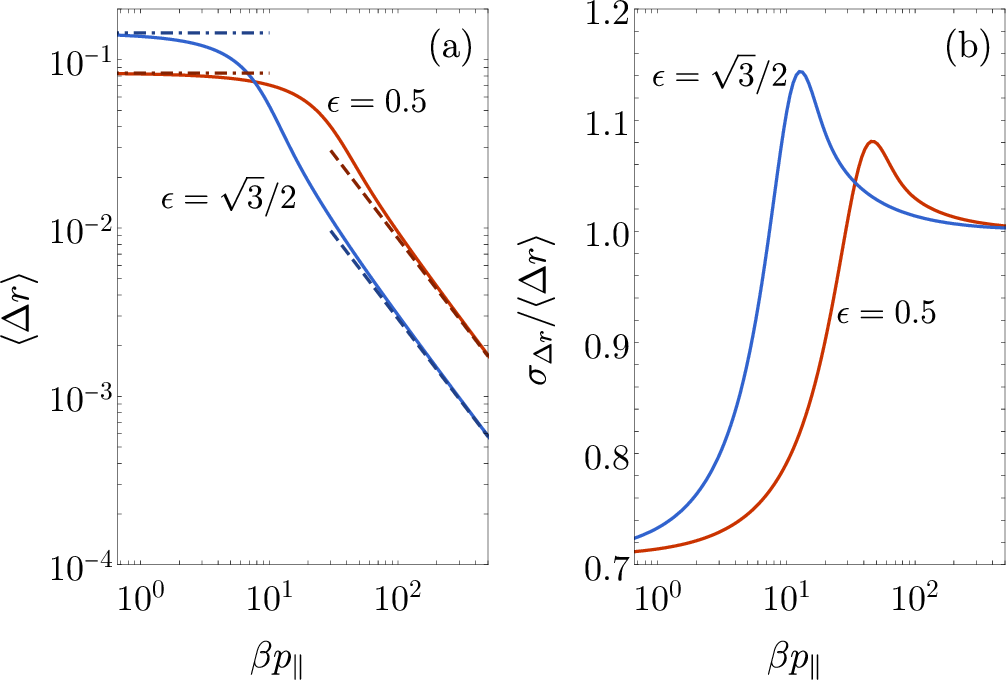}
	\caption{Plot of (a) the average distance $\langle \Delta r \rangle$ and (b) the relative standard deviation $\sigma_{\Delta r}/\langle \Delta r \rangle$  as functions of  the pressure for two values of $\epsilon$. Dashed-dotted and dashed lines in panel (a) represent the low- and high-pressure approximations, respectively.}
	\label{fig:Deltarn}
\end{figure}

Figure~\ref{fig:Deltarn}(a) shows the average radial distance from the cylinder wall, defined in Eq.~\eqref{eq:deltarn} with $n=1$, along with its low- and high-pressure approximations from Eqs.~\eqref{deltarnlowbp} and~\eqref{eq:deltarn_HP}, respectively, for two values of $\epsilon$.
As pressure increases, the average radial position shifts toward the wall from $\langle \Delta r \rangle={\epsilon}/{6}$ (corresponding to a uniform distribution) at low pressure to $\langle \Delta r \rangle\sim 1/\bp$ (corresponding to a distribution concentrated near the wall) at high pressure.

\begin{figure}
	\includegraphics[width=\columnwidth]{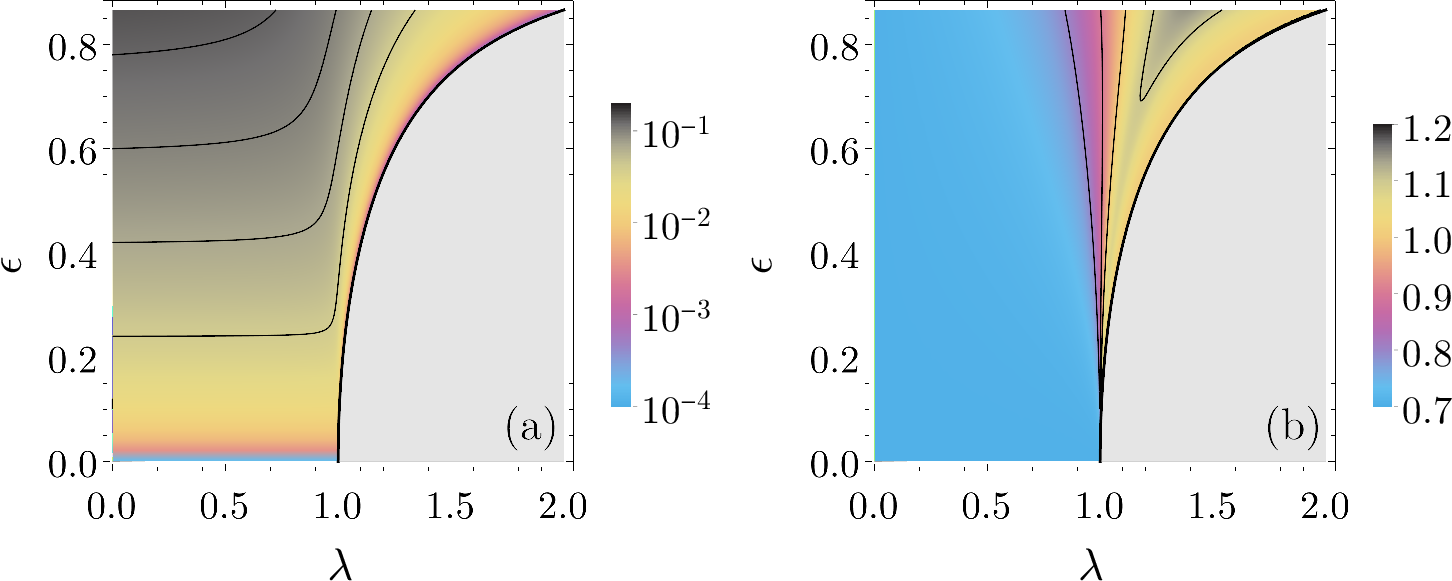}
	\caption{Contour plots of (a) $\langle \Delta r \rangle$ and (b)  $\sigma_{\Delta r}/\langle \Delta r \rangle$ as functions of $\lambda$ and $\epsilon$. In panel (a), the contour lines correspond to the values $\langle \Delta r \rangle=0.13$, $0.10$, $0.07$, and $0.04$, from top to bottom. In panel (b), the contour lines correspond to  $\sigma_{\Delta r}/\langle \Delta r \rangle=0.8$, $0.9$, $1.0$, and $1.1$, from left to right.}
	\label{fig:Delta_contour}
\end{figure}

The positional fluctuations around the average position are quantified by the standard deviation $\sigma_{\Delta r}$, as defined in Eq.~\eqref{sigma_r}. Its value, relative to the mean displacement $\langle \Delta r \rangle$, is shown in Fig.~\ref{fig:Deltarn}(b). The ratio $\sigma_{\Delta r} / \langle \Delta r \rangle$ approaches $1/\sqrt{2}$ and $1$ in the low- and high-pressure limits, respectively, regardless of the excess pore diameter $\epsilon$. Interestingly, its dependence on pressure is nonmonotonic and displays a maximum that becomes increasingly sharp as $\epsilon$ increases. All of these features are also evident in the contour maps in Fig.~\ref{fig:Delta_contour}.

\begin{figure}
	\includegraphics[width=0.95\columnwidth]{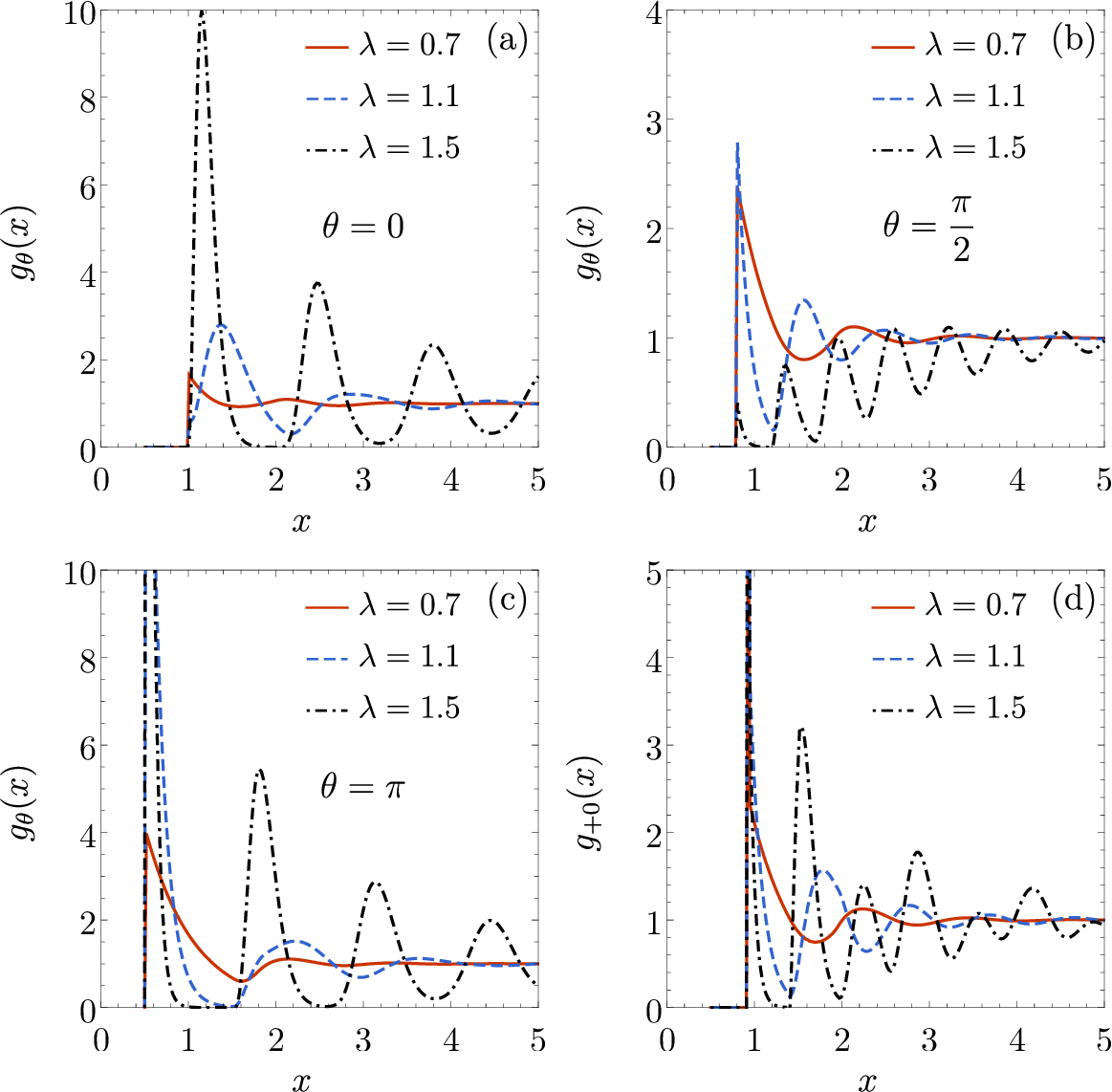}
	\caption{Plot of the longitudinal RDFs $g_\theta(x)$, as defined in Eq.~\eqref{gtheta}, with  (a) $\theta=0$, (b) $\theta={\pi}/{2}$, (c) $\theta=\pi$, and (d) $g_{+0}(x)$ for $\epsilon=\sqrt{3}/2$ and three values of the linear density: $\lambda=0.7$, $1.1$, and $1.5$. The contact distance values in panels (a)--(d) are $1$, $\sqrt{{5}/{8}}\simeq 0.79$, $0.5$, and ${\sqrt{13}}/{4}\simeq 0.90$, respectively.}
	\label{fig:GxPartials}
\end{figure}

\subsection{Longitudinal partial radial distribution functions}

The RDF, which measures spatial correlations between particles, is a key quantity for understanding the ordering of particles. The method described in Sec.~\ref{sec:struct} allows us to obtain not only the \emph{total} longitudinal RDF $g(x)$ defined in Eq.~\eqref{eq:gtotal}, but also the \emph{partial} correlation functions $g_{\rr_1,\rr_2}(x)$ defined in Eq.~\eqref{eq:gpartial}, which account for spatial correlations between particles at specific transverse positions $\rr_1$ and $\rr_2$.

At high pressure, particles tend to accumulate near the wall to achieve the close-packing structure. Therefore, the most relevant partial correlation functions are those of peripheral particles, i.e.,
\beq
\label{gtheta}
g_{\theta}(x)\equiv \left.g_{\rr_1,\rr_2}(x)\right|_{r_1=r_2=\frac{\epsilon}{2}},\quad \theta=\theta_{12}.
\eeq
Similarly,
\beq
g_{+0}(x)\equiv \left.g_{\rr_1,\rr_2}(x)\right|_{r_1=\frac{\epsilon}{2},r_2=0}
\eeq
characterizes the spatial correlations between a peripheral particle and another one on the cylinder axis.

Figure~\ref{fig:GxPartials} presents the partial functions  $g_{\theta}(x)$ with $\theta=0,\pi,{\pi}/{2}$, as well as $g_{+0}(x)$, for $\epsilon=\sqrt{3}/2$ at three different densities. Apart from the fact that each RDF becomes nonzero only after the corresponding contact distance $a_{\rr_1,\rr_2}$, they behave quite differently from each other, especially at higher densities, where the zigzag structure starts developing.

\begin{figure}
	\includegraphics[width=\columnwidth]{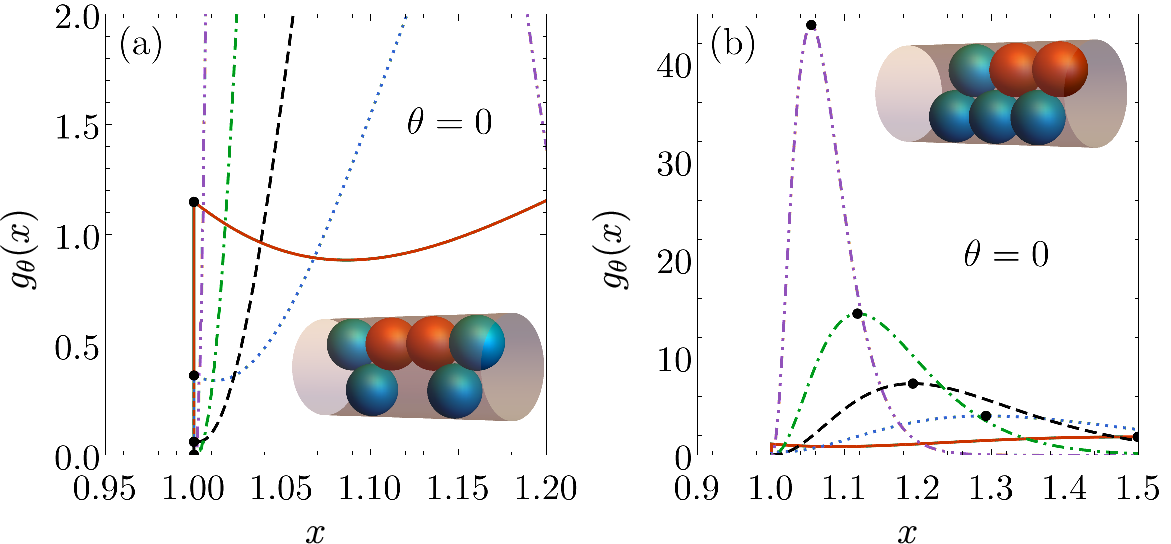}
	\caption{Plot of $g_{\theta}(x)$ with $\theta=0$ near $x=1$ for $\epsilon=\sqrt{3}/2$ at densities $\lambda=1$ (solid line), $\lambda=1.2$ (dotted line), $\lambda=1.4$ (dashed line), $\lambda=1.6$ (dashed-dotted line), and $\lambda=1.8$ (dashed-double-dotted line). Circles represent the local maxima associated with (a) first and (b) second nearest neighbors. Insets show the characteristic particle arrangements for each case.}
	\label{fig:GxPlusPlus}
\end{figure}

At $\lambda=1.5$, correlations between peripheral particles, as shown in Figs.~\ref{fig:GxPartials}(a)--(c), exhibit a distinct solid-like structure characterized by well-defined, ordered minima and maxima. In contrast, $g_{+0}(x)$ retains a more liquid-like behavior, lacking the pronounced ordering observed in the peripheral correlations.

In the case of $g_{0}(x)$, Fig.~\ref{fig:GxPartials}(a) shows that the value at contact ($x=a_{\rr_1,\rr_2}=1$) decreases with increasing density until this peak is no longer noticeable. In fact, the first peak visible at $\lambda=1.5$ in Fig.~\ref{fig:GxPartials}(a) corresponds to the second nearest-neighbor contribution at a longitudinal distance slightly larger than $a^\two_{\rr_1,\rr_2}=1$ [see Eq.~\eqref{atwo}].
The behavior of the peak position and height of $g_{0}(x)$ for the first and second nearest neighbors is tracked in Figs.~\ref{fig:GxPlusPlus}(a) and (b), respectively, for different densities. As density increases, the occurrence of a ``defect'' consisting of two first nearest neighbors with the same orientation ($\theta=0$) is strongly suppressed [see the inset in Fig.~\ref{fig:GxPlusPlus}(a)], while the opposite occurs for two second nearest neighbors [see the inset in Fig.~\ref{fig:GxPlusPlus}(b)].

In contrast to $g_{0}(x)$, the contact value of $g_{\pi}(x)$ (at $x=a_{\rr_1,\rr_2}=0.5$) increases rapidly with increasing density [see Fig.~\ref{fig:GxPartials}(c)], as expected from the formation of zigzag configurations.
The most peculiar behavior is observed in $g_{{\pi}/{2}}(x)$ [see Fig.~\ref{fig:GxPartials}(b)], where the values of the RDF and its oscillations for the first few neighbors decrease with increasing density. This is because, at high pressure, the first nearest neighbor of a peripheral particle tends to minimize the longitudinal separation by positioning itself at a relative angle near $\theta=\pi$, while the second nearest neighbor tends to occupy an angle near $\theta=0$. In this structure, the relative angle $\theta=\pi/2$ is unfavorable for any of the first few nearest neighbors, leading to a decrease in the peaks of $g_{{\pi}/{2}}(x)$ with increasing pressure. However, for sufficiently large $x$, this effect becomes progressively blurred, and the expected limit $\lim_{x\to\infty}g_{{\pi}/{2}}(x)=1$ is reached.

\begin{figure}
	\includegraphics[width=\columnwidth]{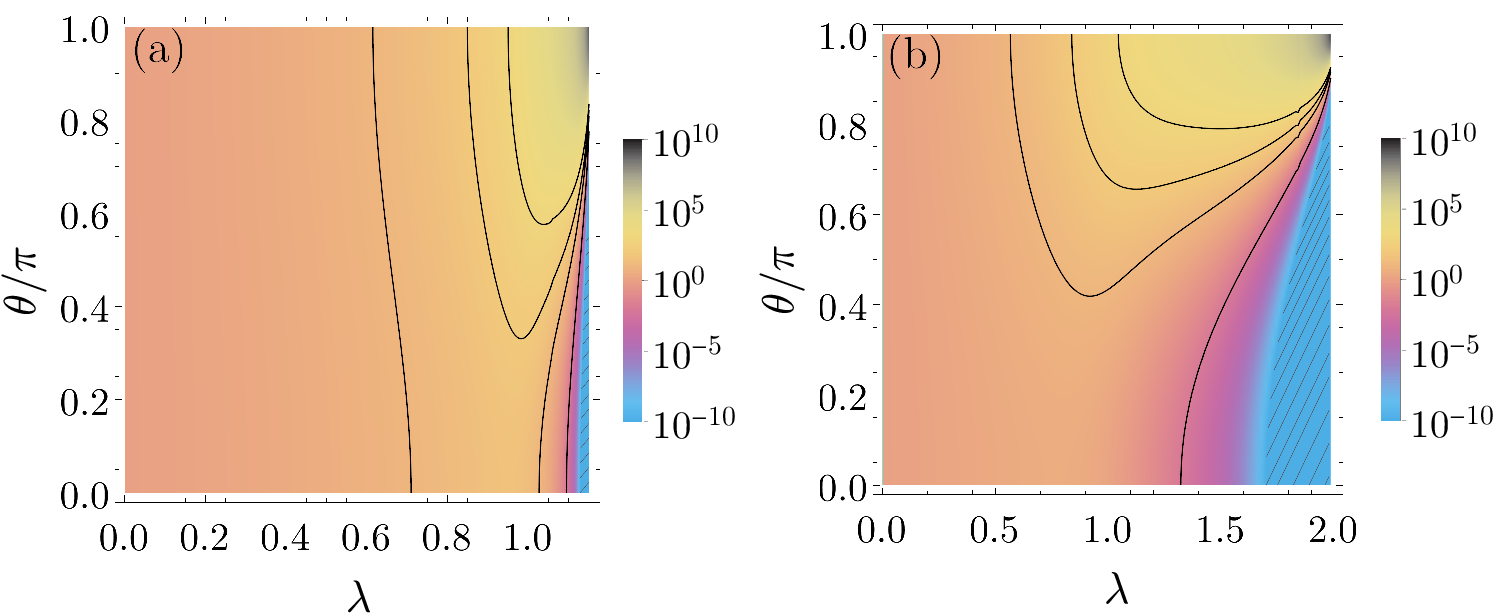}
	\caption{Contour plots of $g_\theta^\cont$ as a function of $\lambda$ and $\theta$ for (a) $\epsilon = 0.5$ and (b) $\epsilon = \sqrt{3}/2$. In each panel, the hatched regions indicate parameter ranges where $g_\theta^\cont < 10^{-10}$. The contour lines correspond to the values $g_\theta^\cont = 10^{-2}$, $10$, $10^2$, and $10^3$, with values increasing as the lines move away from the hatched regions. Note that a left branch of the $g_\theta^\cont = 10$ contour line appears in panel (a).}
	\label{fig:ContourGcontact}
\end{figure}

We now analyze the high-pressure behavior of the contact values of $g_\theta(x)$. By inserting Eq.~\eqref{eq:phiell_HP} into Eq.~\eqref{contact}, one finds
\begin{align}\label{eq:contactpeak}
g_\theta^\cont=&\sqrt{\frac{\pi}{2}}\epsilon(1-\epsilon^2)^{1/4}(\bp)^{3/2}
\nn&\times
e^{-\bp\left(\sqrt{1-\epsilon^2\sin^2\frac{\theta}{2}}-\sqrt{1-\epsilon^2}\right)}.
\end{align}
This contact value decays quasi-exponentially with increasing pressure if $\theta\neq \pi$, with faster decay as $\theta$ evolves from $\pi$ to $0$.
In the special case $\theta=\pi$, however, the contact value increases algebraically as $\sim (\bp)^{3/2}$.

Figure~\ref{fig:ContourGcontact} shows contour plots of $g_\theta^\cont$ for (a) $\epsilon = 0.5$ and (b) $\epsilon = \sqrt{3}/2$. Near close packing, the contact value for peripheral spheres increases by several orders of magnitude as the relative orientation $\theta$ changes from $0$ to $\pi$, with this effect becoming more pronounced as $\epsilon$ increases. In addition, for a fixed value of $\theta < \pi$, $g_\theta^\cont$ displays a nonmonotonic dependence on $\lambda$.

\begin{figure}
	\includegraphics[width=0.95\columnwidth]{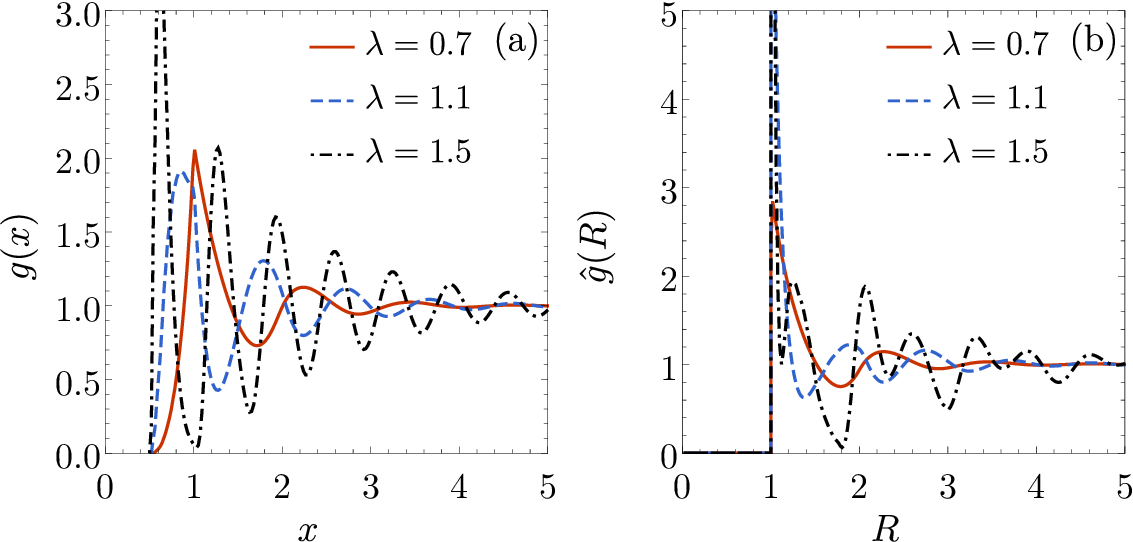}
	\caption{Plot of (a) $g(x)$ and (b) $\hat{g}(R)$ for $\epsilon=\sqrt{3}/2$ at three different densities.}
	\label{fig:GxAndGr}
\end{figure}

\begin{figure}
	\includegraphics[width=0.95\columnwidth]{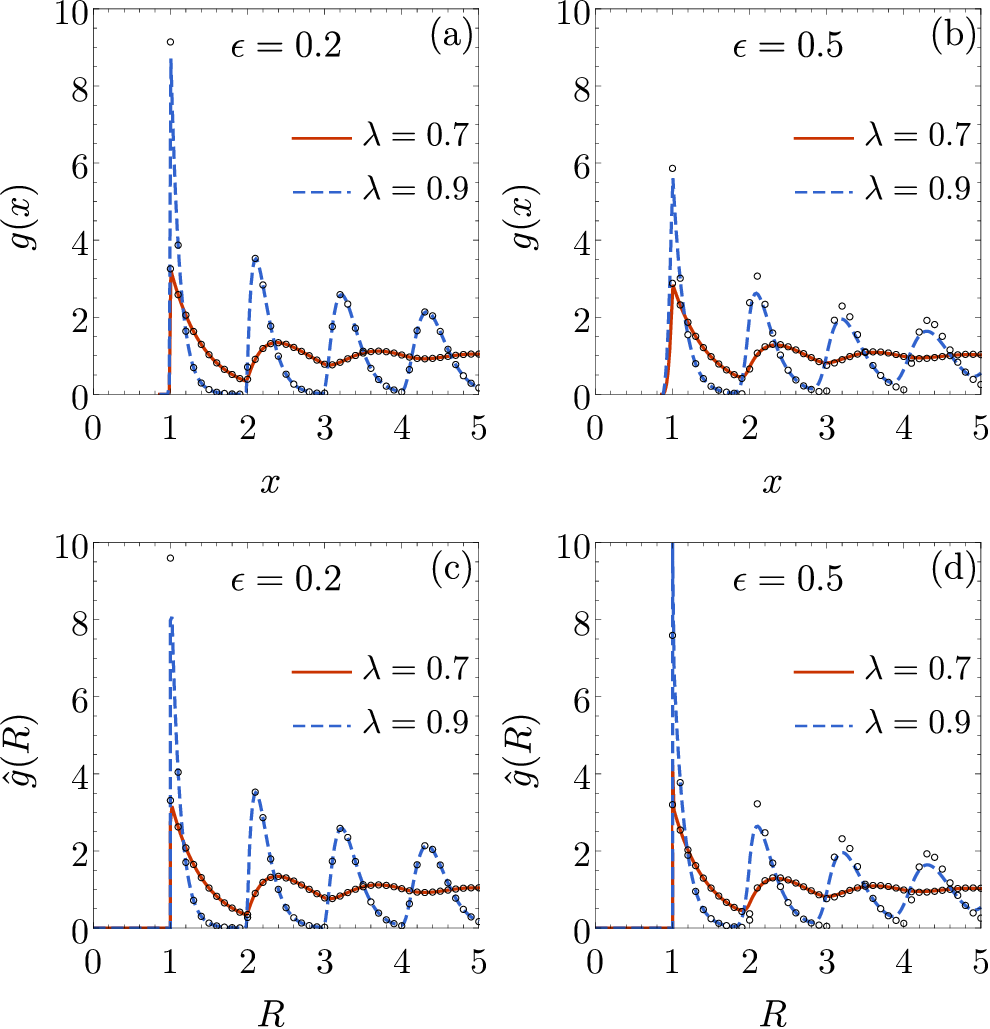}
	\caption{Plot of (a) and (b) $g(x)$ and (c) and (d) $\hat{g}(R)$ at densities $\lambda=0.7,0.9$  and for two values of the pore size: (a) and (c) $\epsilon=0.2$ and (b) and (d) $\epsilon=0.5$. Solid and dashed lines represent the exact curves whereas the open circles represent the corresponding small-$\epsilon$ approximation.}
	\label{fig:GxEpsSmall}
\end{figure}

\subsection{Total radial distribution functions}

Let us now examine the spatial correlation functions between all particles, irrespective of their transverse position. Figure~\ref{fig:GxAndGr} presents both the longitudinal RDF $g(x)$ [Eq.~\eqref{eq:gtotal}] and the nominal RDF $\hat{g}(R)$ [Eq.~\eqref{eq:nhat}] at several densities. The oscillatory behavior in $g(x)$ emerges at lower densities than in $\hat{g}(R)$, but the latter exhibits greater complexity in the positioning of local maxima at $\lambda=1.5$, similar to the case of confined hard disks.\cite{MS24b}
A key distinction between the two functions is the shift in the position of the first peak in $g(x)$, which moves from $x \simeq 1$ at $\lambda=0.7$ to $x \simeq 0.5$ at $\lambda=1.5$, reflecting the emergence of zigzag ordering. In contrast, the first peak in $\hat{g}(R)$ remains fixed at $R = 1$. If $x_n$ and $R_n$ denote the locations of the first few peaks of $g(x)$ and $\hat{g}(R)$, respectively, the zigzag ordering manifests in the high-pressure trends $R_n \simeq \sqrt{x_n^2+\epsilon^2}$ for odd $n$ and $R_n \simeq x_n$ for even $n$.

The evaluation of  $g(x)$ and $\hat{g}(R)$ is computationally expensive due to the double integrals in Eqs.~\eqref{eq:gtotal} and~\eqref{eq:nhat}. It is, therefore, useful to assess the accuracy of the expansions in powers of $\epsilon$ from Eqs.~\eqref{eq:Gs_SEnew} and \eqref{Gnew} for different pore sizes, as these provide an efficient approximate method for evaluating both RDFs.

Figure~\ref{fig:GxEpsSmall} shows the comparison of the approximation with the exact solution for $g(x)$ and $\hat{g}(R)$. In applying the approximations from Eqs.~\eqref{eq:Gs_SEnew} and \eqref{Gnew}, we retained the exact equation of state rather than using the approximate form in Eq.~\eqref{eq:zpar_smalle}.

For a small pore size parameter ($\epsilon = 0.2$), the approximation remains highly accurate across a broad range of densities. When $\epsilon = 0.5$, it continues to perform well at low and moderate densities (e.g., $\lambda = 0.7$), but noticeable deviations appear at higher densities (e.g., $\lambda = 0.9$). For small values of $\epsilon$, the curves $g(x)$ and $\hat{g}(R)$ are nearly indistinguishable since the distance $R$ between two particles closely matches their longitudinal separation $x$. This similarity fades as $\epsilon$ increases, as seen by comparing Fig.~\ref{fig:GxAndGr}(a) with Fig.~\ref{fig:GxAndGr}(b) and Fig.~\ref{fig:GxEpsSmall}(b) with Fig.~\ref{fig:GxEpsSmall}(d).

\section{Conclusions}\label{sec:conclusions}

In this work, we extended the mapping method originally developed for Q1D hard disks to derive the exact anisotropic thermodynamic and structural properties of hard spheres confined within a cylindrical pore. The theory was adapted to account for the additional degree of freedom in the confined directions, and numerical techniques were developed to compute relevant quantities with high accuracy.

For thermodynamic properties, we recovered the longitudinal equation of state previously obtained via the transfer-matrix method and additionally computed the transverse component. A crossover in the anisotropic pressure components was identified: at sufficiently high densities, the transverse compressibility factor $Z_{\perp}$ exceeds the longitudinal one $Z_{\|}$ when the pore width surpasses a critical threshold, $\epsilon = \sqrt{2/3}$.

We also derived analytical expressions in the limit of small pore sizes, where the system approaches the Tonks gas. In addition, for a fixed pore width, we obtained both low- and high-pressure asymptotics, with the low-pressure limit yielding the second and third virial coefficients for both longitudinal and transverse pressures.

Regarding structural properties, we computed the longitudinal RDF $g(x)$ and the 3D RDF-like function $\hat{g}(R)$, analyzing how particle ordering along the pore evolves with increasing density. Using the longitudinal partial RDF at specific transverse positions, we quantified the disappearance of defects near close packing, finding that it follows a $(\bp)^{3/2}\exp[-\bp(1-\sqrt{1-\epsilon^2})]$ pressure dependence.

It is worth noting that the planar zigzag arrangement formed by identical spheres near close packing in cylindrical confinement share a geometrical equivalence with the zigzag structure of identical disks in parallel-slit confinement.\cite{GV13,MS23b,V25,F91} In both cases, the projection of the particle centers onto the longitudinal direction leads to the same underlying geometry, governed by the same contact condition for nearest neighbors. Consequently, in the close-packing limit, the maximum linear density is the same as in the corresponding two-dimensional system of hard disks in narrow slit pores. However, the thermodynamic and structural quantities, such as pressure components and radial distribution functions, are specific to the 3D cylindrical geometry due to the distinct confinement topology and accessible configuration space. Therefore, while our results reveal a broader relevance for characterizing zigzag ordering in Q1D systems, they should be interpreted with this geometrical and dimensional mapping in mind.

The theoretical framework developed here can also be extended to systems where particles interact with the confining walls through more than just hard-core exclusion via attractive or repulsive interactions near the walls. Such extensions are especially relevant in experimental contexts, where wall-particle interactions are often significant. Another natural direction involves introducing interparticle forces beyond the hard-sphere model, as we previously did in the case of confined hard disks with attractive or repulsive coronas.\cite{MS24} Continuous interaction tails, such as Yukawa-like potentials, can also be treated exactly, provided the potential is truncated to ensure interactions remain limited to first nearest neighbors.

To conclude, the results presented in this work provide a rigorous and versatile framework for understanding the interplay between confinement and ordering in Q1D fluids. The analytical and numerical methods developed here can be extended to explore other cross-sectional geometries, interaction models, or external fields, offering new insight into the behavior of confined fluids in both nanoscale and biological settings.

\acknowledgments
Financial support from Grant No.~PID2020-112936GB-I00, funded by MCIN/AEI/10.13039/501100011033
is gratefully acknowledged.
A.M.M. is grateful to the Spanish Ministerio de Ciencia e Innovaci\'on for a predoctoral fellowship PRE2021-097702.

\section*{AUTHOR DECLARATIONS}
\subsection*{Conflict of Interest}
The authors have no conflicts to disclose.
\subsection*{Author Contributions}
\textbf{Ana M. Montero}: Formal analysis (equal); Investigation (equal);
Methodology (equal); Software (lead); Writing -- original draft
(lead). \textbf{Andr\'es Santos}: Conceptualization (lead); Formal analysis
(equal); Funding acquisition (lead); Investigation (equal); Methodology
(equal); Supervision (lead); Writing -- original draft (supporting);
Writing -- review \& editing (lead).

\section*{Data availability}
The data that support the findings of this study are available
from the corresponding author upon reasonable request.

\appendix

\section{Proof of the contact theorem, Eq.~\eqref{eq:cont_th}}
\label{appA}
To prove Eq.~\eqref{eq:cont_th}, let us differentiate with respect to $u_1$ on both sides
of Eq.~\eqref{eq:eigenfunction2} and then multiply by $u_1\fuo$. This yields
\bal
u_1\frac{\partial \phi^2_{u_1}}{\partial u_1}=&\frac{\bp}{2\ell}\fuo\intu du_2\,\fup\intt d\theta_{12}e^{-a_{\rr_1,\rr_2}\bp}\nn
&\times \frac{u_1-\sqrt{u_1 u_2}\cos\theta_{12}}{a_{\rr_1,\rr_2}}.
\eal
Next, we integrate over $u_1$,
\bal
\label{E.21}
\intu du_1\,u_1\frac{\partial \phi^2_{u_1}}{\partial u_1}=&\frac{\bp}{4\ell}\intu du_1\,\fuo\intu du_2\,\fup\intt d\theta_{12}\,e^{-a_{\rr_1,\rr_2}\bp}\nn
&\times \frac{1-a_{\rr_1,\rr_2}^2}{a_{\rr_1,\rr_2}}.
\eal
To obtain the right-hand side, first, we have made the exchange $u_1\leftrightarrow u_2$ inside the double integral and then we have taken the arithmetic mean of both expressions.
Integrating by
parts, the left-hand side of Eq.~\eqref{E.21} gives $(\epsilon^2/4)\phi^2_{\epsilon^2/4}-1/\pi$, while the right-hand side can be recognized as
$(Z_\perp-1)/\pi$ in view of Eq.~\eqref{eq:zperp}. This proves Eq.~\eqref{eq:cont_th}.

\section{\label{app:mathematicalaspects}Mathematical aspects of limiting behaviors}
\subsection{\label{app:smallepsapp}Small $\epsilon$}
By performing the change of variable from Eq.~\eqref{eq:resc} on Eqs.~\eqref{eq:normalization_1} and \eqref{eq:eigenfunction2}, one obtains
\begin{subequations}
	\beq
	\label{eq:resc1}
	\pi\int_0^{\frac{1}{4}} d\overline{u}\,\overline{\phi}^2_{\overline{u}}=1,
	\eeq
	\beq
	\label{eq:resc2}
	\frac{1}{2}\int_0^{\frac{1}{4}} d\overline{u}_2\,\overline{\phi}_{\overline{u}_2}\int_0^{2\pi}d\td\,e^{-a_{\rr_1,\rr_2}\bp}=\overline{\ell}\overline{\phi}_{\overline{u}_1},
	\eeq
	\end{subequations}
where, in terms of $\overline{u}_1$ and $\overline{u}_2$, the quantity $a_{\rr_1,\rr_2}$ is expressed as
\beq
\label{eq:resc4}
a_{\rr_1,\rr_2}=\sqrt{1-\epsilon^2\left(\overline{u}_1+\overline{u}_2-2\sqrt{\overline{u}_1\overline{u}_2}\cos\td\right)}.
\eeq

Expanding in powers of $\epsilon$, we have
\bal
\label{B3}
e^{-a_{\rr_1,\rr_2}\bp}=&  e^{-\bp}\left[1+\frac{\bp}{2}\epsilon^2\left(\overline{u}_1+\overline{u}_2-2\sqrt{\overline{u}_1\overline{u}_2}\right.\right.\nn
&\left.\left.\times\cos\theta_{12}\right)+\mathcal{O}(\epsilon^4)\right],
\eal
implying the expansions
\begin{subequations}
\label{eq:resc5}
\beq
\overline{\phi}_{\overline{u}}=\frac{2}{\sqrt{\pi}}\left[1+\epsilon^2\overline{\phi}^{(1)}_{\overline{u}}+\mathcal{O}(\epsilon^4)\right],
\eeq
\beq
\overline{\ell}=\frac{\pi}{4}e^{-\bp}\left[1+\epsilon^2 \overline{\ell}^{(1)}+\mathcal{O}(\epsilon^4)\right].
\eeq
\end{subequations}
The normalization condition in Eq.~\eqref{eq:resc1} leads to
\beq\label{eq:resc6}
\int_0^{\frac{1}{4}} d\overline{u}\,\overline{\phi}^{(1)}_{\overline{u}}=0.
\eeq
Taking that into account, inserting Eqs.~\eqref{B3} and \eqref{eq:resc5} into Eq.~\eqref{eq:resc2} gives
\beq
\label{E17}
\frac{\bp}{2}\left(\overline{u}+\frac{1}{8}\right)=\overline{\ell}^{(1)}+\overline{\phi}^{(1)}_{\overline{u}}.
\eeq
This implies that $\overline{\phi}^{(1)}_{\overline{u}}$ equals ${\bp}\overline{u}/2$ plus a term independent of $\overline{u}$, which is determined from Eq.~\eqref{eq:resc6}. The final result is
\beq\label{eq:smalleps0}
\overline{\phi}^{(1)}_{\overline{u}}=\frac{\bp}{2}\left(\overline{u}-\frac{1}{8}\right),\quad \overline{\ell}^{(1)}=\frac{\bp}{8}.
\eeq
Reverting to the original variables gives Eq.~\eqref{eq:phiell_LP}.

To obtain the small $\epsilon$ limiting behavior of $\widetilde{G}_{{\rr}_1,{\rr}_2}(s)$, we  perform the variable changes from Eq.~\eqref{eq:resc} again and write
\begin{equation}
	\widetilde{G}_{{\rr}_1,{\rr}_2}(s) = \widetilde{G}^\hr(s) \left[1+ \epsilon^2 \gamma_{\bar{\rr}_1,\bar{\rr}_2}(s) + \mathcal{O}(\epsilon^4) \right],
\end{equation}
where $\widetilde{G}^\hr(s)$ is defined in Eq.~\eqref{eq:GsHR}, $\gamma_{\bar{\rr}_1,\bar{\rr}_2}(s)$ is a function to be determined, and $\bar{\rr}\equiv \rr/\epsilon$.
Expanding in powers of $\epsilon$ on both sides of Eq.~\eqref{eq:GsLaplace}, and  taking into account Eqs.~\eqref{eq:resc4} and~\eqref{eq:resc5}, we find that $\gamma_{\bar{\rr}_1,\bar{\rr}_2}(s)$ is a linear function of $\overline{u}_1+\overline{u}_2$ and $\sqrt{\overline{u}_1\,\overline{u}_2}\cos\td$. The coefficients are then determined with the result
\begin{align}
\gamma_{\bar{\rr}_1,\bar{\rr}_2}(s)=&\frac{\lambda}{8} s \widetilde{G}^\hr(s)+\frac{s}{2}(\overline{u}_1+\overline{u}_2)\nn
&-[s+(1-e^{-s})\bp]\sqrt{\overline{u}_1\,\overline{u}_2}\cos\td.
\end{align}
This yields Eq.~\eqref{eq:Gs_SE} after returning to the original variables.

\subsection{\label{app:smallbpapp}Small $\bp$}
Application of the normalization condition on both sides of Eq.~\eqref{eq:getB2_0a} leads to
\beq
\label{eq:getB2_1}
\intu du\,\psi^{(1)}_u=0.
\eeq
Next, inserting Eqs.~\eqref{eq:getB2_0}  into  Eq.~\eqref{eq:eigenfunction2}, we get Eq.~\eqref{eq:getB2_3} with
\begin{subequations}
\label{eq:getB2_2}
\beq\label{eq:getB2_2a}
\Psi^\|_{u_1}\equiv \frac{4}{\epsilon^2}\intu du_2\,\Phi^\|_{u_1,u_2},
\eeq
\begin{align}
\Phi^\|_{u_1,u_2}\equiv &\frac{1}{2\pi}\intt d\td\,a_{\rr_1,\rr_2}\nn
=&\frac{1}{\pi}\left[\sqrt{v^+_{u_1,u_2}}E\left(\frac{-4\sqrt{u_1u_2}}{v^+_{u_1,u_2}}\right)\right.
\nn&
\left.+\sqrt{v^-_{u_1,u_2}}E\left(\frac{4\sqrt{u_1u_2}}{v^-_{u_1,u_2}}\right)\right],
\end{align}
\end{subequations}
where $v^\pm_{u_1,u_2}\equiv 1-(\sqrt{u_1}\pm\sqrt{u_2})^2$ and $E(x)$ is the complete elliptic integral of the second kind.
Insertion of Eq.~\eqref{eq:getB2_3} into Eq.~\eqref{eq:getB2_1} allows us to obtain the expression for $B_{2\|}$ shown in Eq.~\eqref{eq:B2}. Then, expanding Eq.~\eqref{eq:zpar} in powers of $\bp$ and making use of Eq.~\eqref{eq:getB2_0}, one obtains Eq.~\eqref{eq:B3} after some algebra.

Analogously, the expansion of Eq.~\eqref{eq:zperp} yields  the coefficients given by Eqs.~\eqref{eq:B2perp} and \eqref{eq:B3perp}, where
\begin{subequations}
\label{eq:get_B2p1}
	\beq \label{eq:get_B2p0}
	\Psi^\perp_{u_1}\equiv \frac{4}{\epsilon^2}\intu du_2\,\Phi^\perp_{u_1,u_2},\quad \Phi^\perp_{u_1,u_2}\equiv\frac{\Phi_{u_1,u_2}-\Phi^\|_{u_1,u_2}}{2},
	\eeq
\begin{align}
	\Phi_{u_1,u_2} \equiv &\frac{1}{2\pi} \intt \frac{d\td}{a_{\rr_1,\rr_2}} \nn
=& \frac{1}{\pi} \left[ \frac{K\left(\frac{-4\sqrt{u_1u_2}}{v_{u_1,u_2}^+}\right)}{\sqrt{v_{u_1,u_2}^+}} + \frac{K\left(\frac{4\sqrt{u_1u_2}}{v_{u_1,u_2}^-}\right)}{\sqrt{v_{u_1,u_2}^-}} \right],
\end{align}
\end{subequations}
$K(x)$ being the complete elliptic integral of the first kind.

\subsection{\label{app:highbpapp}High $\bp$}
Equation~\eqref{eq:phiell_HP} reflects the fact that, at a given value of $u_1$, the integrand on the left-hand side of Eq.~\eqref{eq:eigenfunction2} exhibits a sharp maximum at $u_2={\epsilon^2}/{4}$ and $\td=\pi$, in which case $a_{\rr_1,\rr_2}\to\amin_{u_1}$. It remains to find the normalization constant $\mathcal{N}_0$. More generally, we define
\beq
\mathcal{N}_n=\pi\intu du\, \left( \frac{\epsilon^2}{4}-u\right)^n e^{-2\amin_u\bp}.
\eeq
Since the minimum value of $\amin_u$ occurs at $u={\epsilon^2}/{4}$, we approximate
\beq
\label{amin_HP}
\amin_u\approx \sqrt{1-\epsilon^2}+\frac{\frac{\epsilon^2}{4}-u}{\sqrt{1-\epsilon^2}}.
\eeq
Therefore,
\begin{align}\label{eq:norm_HP2}
\mathcal{N}_n\approx&\pi  e^{-2\sqrt{1-\epsilon^2}\bp}\intu  du\,\left(\frac{\epsilon^2}{4}-u\right)^n e^{-2\frac{\frac{\epsilon^2}{4}-u}{\sqrt{1-\epsilon^2}}\bp}\nn
\approx&\pi  e^{-2\sqrt{1-\epsilon^2}\bp}n!\left(\frac{\sqrt{1-\epsilon^2}}{2\bp}\right)^{n+1}.
\end{align}
In the second step, we have performed the change of variable $t={\epsilon^2}/{4}-u$ and extended the integration limits $\int_0^{{\epsilon^2}/{4}}dt\to\int_0^\infty dt$.

To obtain the eigenvalue $\ell$, we first expand $a_{\rr_1,\rr_2}$ around  $\td=\pi$ as
\beq
\label{eq:01}
a_{\rr_1,\rr_2}\approx \sqrt{1-(\sqrt{u_1}+\sqrt{u_2})^2}+\frac{\epsilon^2}{8\sqrt{1-\epsilon^2}}(\td-\pi)^2,
\eeq
where the coefficient of $(\td-\pi)^2$ has been evaluated at $u_1=u_2={\epsilon^2}/{4}$.
Expanding $[1-(\sqrt{u_1}+\sqrt{u_2})^2]^{1/2}$ around $u_2={\epsilon^2}/{4}$ gives
\beq
\label{eq:02}
\sqrt{1-(\sqrt{u_1}+\sqrt{u_2})^2}\approx  \amin_{u_1}+\frac{\frac{\epsilon^2}{4}-u_2}{\sqrt{1-\epsilon^2}}.
\eeq
Substituting Eqs.~\eqref{eq:phi_HP}, \eqref{amin_HP}, \eqref{eq:01}, and \eqref{eq:02} into Eq.~\eqref{eq:eigenfunction2} gives
\begin{align}\label{eq:ell_HP1}
\ell=&\frac{1}{2\fuo}\intu du_2\,\fup\intt d\td\,e^{-a_{\rr_1,\rr_2}\bp}\nn
\approx&  \frac{e^{-\sqrt{1-\epsilon^2}\bp}}{2}\intu du_2\,e^{-2\frac{\frac{\epsilon^2}{4}-u_2}{\sqrt{1-\epsilon^2}}\bp}\intt d\td\,e^{-\frac{\epsilon^2\bp}{8\sqrt{1-\epsilon^2}}(\td-\pi)^2}.
\end{align}
As before, making the changes of variables $t = {\epsilon^2}/{4} - u$ and $\vartheta = \td - \pi$, and extending the integration limits $\int_0^{{\epsilon^2}/{4}} dt \to \int_0^\infty dt$ and $\int_{-\pi}^{\pi} d\vartheta \to 2\int_0^\infty d\vartheta$, yields Eq.~\eqref{eq:ell_HP}.

The knowledge of the asymptotic form of $\ell$ allows us to obtain that of the excess free energy from Eq.~\eqref{eq:gex},
\beq
\beta g^\ex\approx\sqrt{1-\epsilon^2}\bp+\frac{3}{2}\ln\frac{\pi^{1/3}\epsilon^2\bp}{2\sqrt{1-\epsilon^2}}.
\eeq
Next, using the thermodynamic relations $Z_\|=1+\bp(\partial \beta g^\ex/\partial \bp)_\epsilon$ and $Z_\perp=1-\epsilon^2(\partial \beta g^\ex/\partial \epsilon^2)_{\bp}$, one can directly obtain the results in Eq.~\eqref{eq:z_HP}.

Finally, let us obtain the asymptotic high-pressure behavior of the moments defined in Eq.~\eqref{eq:deltarn}.
By expanding around $u={\epsilon^2}/{4}$, we have
\beq
\frac{\epsilon}{2}-\sqrt{u}\approx \frac{1}{\epsilon}\left(\frac{\epsilon^2}{4}-u\right).
\eeq
Therefore, in the high-pressure regime,
\beq
\langle(\Delta r)^n\rangle\approx \frac{1}{\epsilon^n}\frac{\mathcal{N}_n}{\mathcal{N}_0}.
\eeq
Equation~\eqref{eq:deltarn_HP} follows from the use of Eq.~\eqref{eq:norm_HP2}.

\section{\label{sec:discretization}Numerical details}

When numerically solving the equations shown in Sec.~\ref{sec:theory}, it becomes necessary to discretize the system, i.e., to transform the polydisperse nature of the mapped 1D mixture onto a discrete number of components. In this discrete version of the mapped 1D mixture, each component is labeled by a pair $\ii\equiv (i_u,i_\theta)$, with $i_u=1,2,\ldots,M_u$ and $i_\theta=1,2,\ldots, M_\theta$. This gives
\begin{subequations}
\begin{equation}
u_{i_u}=i_u \Delta u, \quad \Delta u=\frac{\epsilon^2/4}{M_u},
\end{equation}
\begin{equation}
	\theta_{i_\theta}= (i_\theta-1) \Delta\theta,\quad  \Delta\theta=\frac{2\pi}{M_\theta},
\end{equation}
\end{subequations}
which represent the discretization along the radial and angular variables, respectively. The total number of components is then $M = M_u M_\theta$.

Continuing with the discretization process, the continuous function $\fu$ is represented by the discrete set $\{\phi_{i_u};i_u=1,\ldots,M_u\}$, where
\beq
\frac{1}{2}\Delta u\Delta \theta\phi^2_u\to \phi^2_{i_u}.
\eeq
This definition ensures that the correct normalization is preserved   when discretizing Eq.~\eqref{eq:normalization_1} in the following form:
\beq
\label{eq:normalization_D}
\sum_{\ii}\phi^2_{i_u}=1,
\eeq
where the notation $\sum_{\ii}$ means $\sum_{i_u=1}^{M_u}\sum_{i_\theta=1}^{M_\theta}$.
The eigenvalue problem, Eq.~\eqref{eq:eigenfunction2}, becomes
\beq
\label{E.13}
\sum_{\jj}\phi_{j_{u}}e^{-a_{\ii\jj} \bp}=\frac{\bp}{A^2}\phi_{i_{u}},
\eeq
where
\begin{subequations}
	\beq
	A^2=\frac{\bp}{2\ell} \Delta u\Delta\theta,
	\eeq
		\beq
	a_{\ii\jj}=\sqrt{1-\left[u_{i_u} + u_{j_u}-2\sqrt{u_{i_u} u_{j_u}}\cos(\theta_{i_\theta}-\theta_{j_\theta}) \right]}.
	\eeq
\end{subequations}
Analogously, the discrete versions of Eq.~\eqref{eq:zpar&zperp} are
\begin{subequations}
	\beq
	Z_\|=1+A^2  \sum_{\ii,\jj}\phi_{i_u}\phi_{j_u}e^{-a_{\ii\jj}\bp}a_{\ii\jj},
	\eeq
	\beq
	Z_\perp=1+\frac{A^2}{2} \sum_{\ii,\jj}\phi_{i_u}\phi_{j_u}e^{-a_{\ii\jj}\bp} \frac{1-a_{\ii\jj}^2}{a_{\ii\jj}}.
	\eeq
\end{subequations}

Regarding the correlation functions, the discretized versions of Eqs.~\eqref{eq:Omega01}, \eqref{eq:PP1s}, \eqref{eq:Gs01}, and \eqref{eq:gtotalLaplace} are
\begin{subequations}
\beq
\Omega_{\ii\jj}(s)=\frac{e^{-a_{\ii\jj}s}}{s},
\eeq
\beq
\widetilde{\PP}_{\ii\jj}^{(1)}(s)=A^2\frac{\phi_{j_u}}{\phi_{i_u}}\Omega_{\ii\jj}(s+\bp)
\eeq
\beq
\widetilde{G}_{\ii\jj}(s)=
\frac{1}{\lambda\phi_{j_u}^2}\left(\widetilde{\mathsf{P}}^{(1)}(s)\cdot\left[\mathsf{I}-\widetilde{\mathsf{P}}^{(1)}(s)\right]^{-1}\right)_{\ii\jj},
\eeq
\beq
\widetilde{G}(s)=\sum_{\ii,\jj}\phi_{i_u}^2\phi_{j_u}^2\widetilde{G}_{\ii\jj}(s).
\eeq
\end{subequations}

\begin{figure}
	\includegraphics[width=0.5\columnwidth]{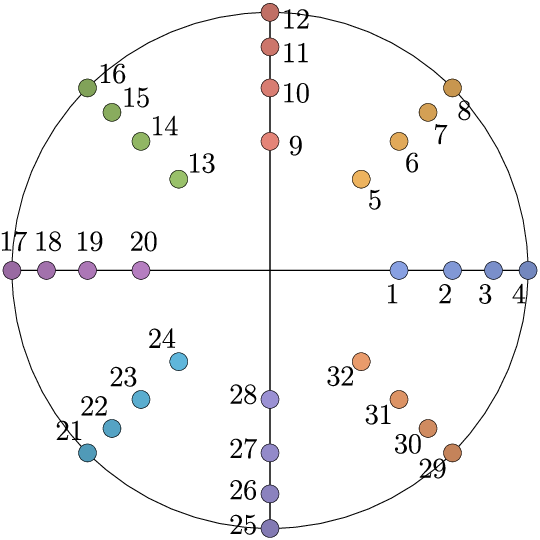}
	\caption{Schematic representation of the discretization of the transverse positions. The shown example corresponds to a labeling scheme used for a system with $M_\theta = 8$ and $M_u=4$. The circles represent the centers of the spheres.}
	\label{fig:Discretization}
\end{figure}

From a practical point of view, it is useful to assign a single label $i=1,\ldots,M$ to each component. Such an assignment is arbitrary, and any permutation is equally valid. However, some permutations are more advantageous than others, as they preserve symmetries that facilitate numerical computations. In particular, the labeling scheme used throughout all calculations is
\begin{equation}
	i = \begin{cases}
		 i_u + (i_\theta - 1)M_u, & 1\leq i_\theta \leq M_\theta/2,\\
		M_u - (i_u - 1) + (i_\theta - 1)M_u, & M_\theta/2<i_\theta\leq M_\theta,
		\end{cases}
\end{equation}
where $M_\theta$ is always assumed to be an even number. An example of this labeling is shown in Fig.~\ref{fig:Discretization}.

The solution to the eigenfunction problem in Eq.~\eqref{eq:eigenfunction2} and the computation of thermodynamic properties in Eq.~\eqref{eq:zpar&zperp} were handled semi-discretely by numerically evaluating integrals of the form $\intt d \theta_{12} \cdots$ and using $M_u\sim 10^3$. However, this approach is no longer valid when dealing with structural properties. In this case, as mentioned earlier, the total number of components in the discrete mixture is $M = M_u M_\theta$, with each factor adjustable independently. Empirical results indicate that increasing $M_u$ is generally more effective in approaching the polydisperse limit than increasing $M_\theta$, meaning that radial discretization plays a more critical role than orientational discretization. Typical values used are $M_u=50$ and $M_\theta= 2^m$ with $m=5$.

To minimize discretization effects, each quantity of interest was computed for several values of the discretization parameters and then extrapolated to the limit $M_u \to \infty$ (and $M_\theta \to \infty$ for structural quantities) by plotting it against $1/M_u$ (and $1/M_\theta$). This approach achieves convergence to the polydisperse limit more efficiently than merely increasing $M_u$ and/or $M_\theta$.

\section*{REFERENCES}


\begin{thebibliography}{78}%
\makeatletter
\providecommand \@ifxundefined [1]{%
 \@ifx{#1\undefined}
}%
\providecommand \@ifnum [1]{%
 \ifnum #1\expandafter \@firstoftwo
 \else \expandafter \@secondoftwo
 \fi
}%
\providecommand \@ifx [1]{%
 \ifx #1\expandafter \@firstoftwo
 \else \expandafter \@secondoftwo
 \fi
}%
\providecommand \natexlab [1]{#1}%
\providecommand \enquote  [1]{``#1''}%
\providecommand \bibnamefont  [1]{#1}%
\providecommand \bibfnamefont [1]{#1}%
\providecommand \citenamefont [1]{#1}%
\providecommand \href@noop [0]{\@secondoftwo}%
\providecommand \href [0]{\begingroup \@sanitize@url \@href}%
\providecommand \@href[1]{\@@startlink{#1}\@@href}%
\providecommand \@@href[1]{\endgroup#1\@@endlink}%
\providecommand \@sanitize@url [0]{\catcode `\\12\catcode `\$12\catcode
  `\&12\catcode `\#12\catcode `\^12\catcode `\_12\catcode `\%12\relax}%
\providecommand \@@startlink[1]{}%
\providecommand \@@endlink[0]{}%
\providecommand \url  [0]{\begingroup\@sanitize@url \@url }%
\providecommand \@url [1]{\endgroup\@href {#1}{\urlprefix }}%
\providecommand \urlprefix  [0]{URL }%
\providecommand \Eprint [0]{\href }%
\providecommand \doibase [0]{https://doi.org/}%
\providecommand \selectlanguage [0]{\@gobble}%
\providecommand \bibinfo  [0]{\@secondoftwo}%
\providecommand \bibfield  [0]{\@secondoftwo}%
\providecommand \translation [1]{[#1]}%
\providecommand \BibitemOpen [0]{}%
\providecommand \bibitemStop [0]{}%
\providecommand \bibitemNoStop [0]{.\EOS\space}%
\providecommand \EOS [0]{\spacefactor3000\relax}%
\providecommand \BibitemShut  [1]{\csname bibitem#1\endcsname}%
\let\auto@bib@innerbib\@empty
\bibitem [{\citenamefont {Pusey}\ and\ \citenamefont {van Megen}(1986)}]{PM86}%
  \BibitemOpen
  \bibfield  {author} {\bibinfo {author} {\bibfnamefont {P.~N.}\ \bibnamefont
  {Pusey}}\ and\ \bibinfo {author} {\bibfnamefont {W.}~\bibnamefont {van
  Megen}},\ }\bibfield  {title} {\enquote {\bibinfo {title} {Phase behaviour of
  concentrated suspensions of nearly hard colloidal spheres},}\ }\href
  {https://doi.org/10.1038/320340a0} {\bibfield  {journal} {\bibinfo  {journal}
  {Nature}\ }\textbf {\bibinfo {volume} {320}},\ \bibinfo {pages} {340--342}
  (\bibinfo {year} {1986})}\BibitemShut {NoStop}%
\bibitem [{\citenamefont {van Megen}\ and\ \citenamefont
  {Underwood}(1994)}]{MU94}%
  \BibitemOpen
  \bibfield  {author} {\bibinfo {author} {\bibfnamefont {W.}~\bibnamefont {van
  Megen}}\ and\ \bibinfo {author} {\bibfnamefont {S.~M.}\ \bibnamefont
  {Underwood}},\ }\bibfield  {title} {\enquote {\bibinfo {title} {Glass
  transition in colloidal hard spheres: {M}easurement and mode-coupling-theory
  analysis of the coherent intermediate scattering function},}\ }\href
  {https://doi.org/10.1103/PhysRevE.49.4206} {\bibfield  {journal} {\bibinfo
  {journal} {Phys. Rev. E}\ }\textbf {\bibinfo {volume} {49}},\ \bibinfo
  {pages} {4206--4220} (\bibinfo {year} {1994})}\BibitemShut {NoStop}%
\bibitem [{\citenamefont {Weeks}\ \emph {et~al.}(2000)\citenamefont {Weeks},
  \citenamefont {Crocker}, \citenamefont {Levitt}, \citenamefont {Schofield},\
  and\ \citenamefont {Weitz}}]{WCLSW00}%
  \BibitemOpen
  \bibfield  {author} {\bibinfo {author} {\bibfnamefont {E.~R.}\ \bibnamefont
  {Weeks}}, \bibinfo {author} {\bibfnamefont {J.~C.}\ \bibnamefont {Crocker}},
  \bibinfo {author} {\bibfnamefont {A.~C.}\ \bibnamefont {Levitt}}, \bibinfo
  {author} {\bibfnamefont {A.}~\bibnamefont {Schofield}},\ and\ \bibinfo
  {author} {\bibfnamefont {D.~A.}\ \bibnamefont {Weitz}},\ }\bibfield  {title}
  {\enquote {\bibinfo {title} {Three-dimensional direct imaging of structural
  relaxation near the colloidal glass transition},}\ }\href
  {https://doi.org/10.1126/science.287.5453.627} {\bibfield  {journal}
  {\bibinfo  {journal} {Science}\ }\textbf {\bibinfo {volume} {287}},\ \bibinfo
  {pages} {627--631} (\bibinfo {year} {2000})}\BibitemShut {NoStop}%
\bibitem [{\citenamefont {Bryant}\ \emph {et~al.}(2002)\citenamefont {Bryant},
  \citenamefont {Williams}, \citenamefont {Qian}, \citenamefont {Snook},
  \citenamefont {Perez},\ and\ \citenamefont {Pincet}}]{BWQSPP02}%
  \BibitemOpen
  \bibfield  {author} {\bibinfo {author} {\bibfnamefont {G.}~\bibnamefont
  {Bryant}}, \bibinfo {author} {\bibfnamefont {S.~R.}\ \bibnamefont
  {Williams}}, \bibinfo {author} {\bibfnamefont {L.}~\bibnamefont {Qian}},
  \bibinfo {author} {\bibfnamefont {I.~K.}\ \bibnamefont {Snook}}, \bibinfo
  {author} {\bibfnamefont {E.}~\bibnamefont {Perez}},\ and\ \bibinfo {author}
  {\bibfnamefont {F.}~\bibnamefont {Pincet}},\ }\bibfield  {title} {\enquote
  {\bibinfo {title} {How hard is a colloidal ``hard-sphere'' interaction?}}\
  }\href {https://doi.org/10.1103/PhysRevE.66.060501} {\bibfield  {journal}
  {\bibinfo  {journal} {Phys. Rev. E}\ }\textbf {\bibinfo {volume} {66}},\
  \bibinfo {pages} {060501} (\bibinfo {year} {2002})}\BibitemShut {NoStop}%
\bibitem [{\citenamefont {Royall}\ \emph {et~al.}(2024)\citenamefont {Royall},
  \citenamefont {Charbonneau}, \citenamefont {Dijkstra}, \citenamefont {Russo},
  \citenamefont {Smallenburg}, \citenamefont {Speck},\ and\ \citenamefont
  {Valeriani}}]{RCDRSSV24}%
  \BibitemOpen
  \bibfield  {author} {\bibinfo {author} {\bibfnamefont {C.~P.}\ \bibnamefont
  {Royall}}, \bibinfo {author} {\bibfnamefont {P.}~\bibnamefont {Charbonneau}},
  \bibinfo {author} {\bibfnamefont {M.}~\bibnamefont {Dijkstra}}, \bibinfo
  {author} {\bibfnamefont {J.}~\bibnamefont {Russo}}, \bibinfo {author}
  {\bibfnamefont {F.}~\bibnamefont {Smallenburg}}, \bibinfo {author}
  {\bibfnamefont {T.}~\bibnamefont {Speck}},\ and\ \bibinfo {author}
  {\bibfnamefont {C.}~\bibnamefont {Valeriani}},\ }\bibfield  {title} {\enquote
  {\bibinfo {title} {Colloidal hard spheres: {T}riumphs, challenges, and
  mysteries},}\ }\href {https://doi.org/10.1103/RevModPhys.96.045003}
  {\bibfield  {journal} {\bibinfo  {journal} {Rev. Mod. Phys.}\ }\textbf
  {\bibinfo {volume} {96}},\ \bibinfo {pages} {045003} (\bibinfo {year}
  {2024})}\BibitemShut {NoStop}%
\bibitem [{\citenamefont {Schmidt}\ and\ \citenamefont {L\"owen}(1996)}]{SL96}%
  \BibitemOpen
  \bibfield  {author} {\bibinfo {author} {\bibfnamefont {M.}~\bibnamefont
  {Schmidt}}\ and\ \bibinfo {author} {\bibfnamefont {H.}~\bibnamefont
  {L\"owen}},\ }\bibfield  {title} {\enquote {\bibinfo {title} {Freezing
  between two and three dimensions},}\ }\href
  {https://doi.org/10.1103/PhysRevLett.76.4552} {\bibfield  {journal} {\bibinfo
   {journal} {Phys. Rev. Lett.}\ }\textbf {\bibinfo {volume} {76}},\ \bibinfo
  {pages} {4552--4555} (\bibinfo {year} {1996})}\BibitemShut {NoStop}%
\bibitem [{\citenamefont {Henderson}, \citenamefont {Sokolowski},\ and\
  \citenamefont {Wasan}(1997)}]{HSW97}%
  \BibitemOpen
  \bibfield  {author} {\bibinfo {author} {\bibfnamefont {D.}~\bibnamefont
  {Henderson}}, \bibinfo {author} {\bibfnamefont {S.}~\bibnamefont
  {Sokolowski}},\ and\ \bibinfo {author} {\bibfnamefont {D.}~\bibnamefont
  {Wasan}},\ }\bibfield  {title} {\enquote {\bibinfo {title} {{Second-order
  Percus-Yevick theory for a confined hard-sphere fluid}},}\ }\href
  {https://doi.org/10.1007/BF02770763} {\bibfield  {journal} {\bibinfo
  {journal} {J. Stat. Phys.}\ }\textbf {\bibinfo {volume} {89}},\ \bibinfo
  {pages} {233--247} (\bibinfo {year} {1997})}\BibitemShut {NoStop}%
\bibitem [{\citenamefont {Fortini}\ and\ \citenamefont
  {Dijkstra}(2006)}]{FD06}%
  \BibitemOpen
  \bibfield  {author} {\bibinfo {author} {\bibfnamefont {A.}~\bibnamefont
  {Fortini}}\ and\ \bibinfo {author} {\bibfnamefont {M.}~\bibnamefont
  {Dijkstra}},\ }\bibfield  {title} {\enquote {\bibinfo {title} {Phase
  behaviour of hard spheres confined between parallel hard plates: manipulation
  of colloidal crystal structures by confinement},}\ }\href
  {https://doi.org/10.1088/0953-8984/18/28/L02} {\bibfield  {journal} {\bibinfo
   {journal} {J. Phys.: Condens. Matter}\ }\textbf {\bibinfo {volume} {18}},\
  \bibinfo {pages} {L371} (\bibinfo {year} {2006})}\BibitemShut {NoStop}%
\bibitem [{\citenamefont {Mittal}, \citenamefont {Errington},\ and\
  \citenamefont {Truskett}(2006)}]{MET06}%
  \BibitemOpen
  \bibfield  {author} {\bibinfo {author} {\bibfnamefont {J.}~\bibnamefont
  {Mittal}}, \bibinfo {author} {\bibfnamefont {J.~R.}\ \bibnamefont
  {Errington}},\ and\ \bibinfo {author} {\bibfnamefont {T.~M.}\ \bibnamefont
  {Truskett}},\ }\bibfield  {title} {\enquote {\bibinfo {title} {Thermodynamics
  predicts how confinement modifies the dynamics of the equilibrium hard-sphere
  fluid},}\ }\href {https://doi.org/10.1103/PhysRevLett.96.177804} {\bibfield
  {journal} {\bibinfo  {journal} {Phys. Rev. Lett.}\ }\textbf {\bibinfo
  {volume} {96}},\ \bibinfo {pages} {177804} (\bibinfo {year}
  {2006})}\BibitemShut {NoStop}%
\bibitem [{\citenamefont {Mittal}, \citenamefont {Errington},\ and\
  \citenamefont {Truskett}(2007)}]{MET07}%
  \BibitemOpen
  \bibfield  {author} {\bibinfo {author} {\bibfnamefont {J.}~\bibnamefont
  {Mittal}}, \bibinfo {author} {\bibfnamefont {J.~R.}\ \bibnamefont
  {Errington}},\ and\ \bibinfo {author} {\bibfnamefont {T.~M.}\ \bibnamefont
  {Truskett}},\ }\bibfield  {title} {\enquote {\bibinfo {title} {Does confining
  the hard-sphere fluid between hard walls change its average properties?}}\
  }\href {https://doi.org/10.1063/1.2748045} {\bibfield  {journal} {\bibinfo
  {journal} {J. Chem. Phys.}\ }\textbf {\bibinfo {volume} {126}},\ \bibinfo
  {pages} {244708} (\bibinfo {year} {2007})}\BibitemShut {NoStop}%
\bibitem [{\citenamefont {Nyg{\aa}rd}, \citenamefont {Sarman},\ and\
  \citenamefont {Kjellander}(2013)}]{NSK13}%
  \BibitemOpen
  \bibfield  {author} {\bibinfo {author} {\bibfnamefont {K.}~\bibnamefont
  {Nyg{\aa}rd}}, \bibinfo {author} {\bibfnamefont {S.}~\bibnamefont {Sarman}},\
  and\ \bibinfo {author} {\bibfnamefont {R.}~\bibnamefont {Kjellander}},\
  }\bibfield  {title} {\enquote {\bibinfo {title} {Local order variations in
  confined hard-sphere fluids},}\ }\href {https://doi.org/10.1063/1.4825176}
  {\bibfield  {journal} {\bibinfo  {journal} {J. Chem. Phys.}\ }\textbf
  {\bibinfo {volume} {139}},\ \bibinfo {pages} {164701} (\bibinfo {year}
  {2013})}\BibitemShut {NoStop}%
\bibitem [{\citenamefont {Thorneywork}\ \emph {et~al.}(2018)\citenamefont
  {Thorneywork}, \citenamefont {Schnyder}, \citenamefont {Aarts}, \citenamefont
  {Horbach}, \citenamefont {Roth},\ and\ \citenamefont {Dullens}}]{TSAHRD18}%
  \BibitemOpen
  \bibfield  {author} {\bibinfo {author} {\bibfnamefont {A.~L.}\ \bibnamefont
  {Thorneywork}}, \bibinfo {author} {\bibfnamefont {S.~K.}\ \bibnamefont
  {Schnyder}}, \bibinfo {author} {\bibfnamefont {D.~G. A.~L.}\ \bibnamefont
  {Aarts}}, \bibinfo {author} {\bibfnamefont {J.}~\bibnamefont {Horbach}},
  \bibinfo {author} {\bibfnamefont {R.}~\bibnamefont {Roth}},\ and\ \bibinfo
  {author} {\bibfnamefont {R.~P.~A.}\ \bibnamefont {Dullens}},\ }\bibfield
  {title} {\enquote {\bibinfo {title} {Structure factors in a two-dimensional
  binary colloidal hard sphere system},}\ }\href
  {https://doi.org/10.1080/00268976.2018.1492745} {\bibfield  {journal}
  {\bibinfo  {journal} {Mol. Phys.}\ }\textbf {\bibinfo {volume} {116}},\
  \bibinfo {pages} {3245--3257} (\bibinfo {year} {2018})}\BibitemShut {NoStop}%
\bibitem [{\citenamefont {Nyg{\aa}rd}\ \emph {et~al.}(2012)\citenamefont
  {Nyg{\aa}rd}, \citenamefont {Kjellander}, \citenamefont {Sarman},
  \citenamefont {Chodankar}, \citenamefont {Perret}, \citenamefont
  {Buitenhuis},\ and\ \citenamefont {van~der Veen}}]{NKSCPBV12}%
  \BibitemOpen
  \bibfield  {author} {\bibinfo {author} {\bibfnamefont {K.}~\bibnamefont
  {Nyg{\aa}rd}}, \bibinfo {author} {\bibfnamefont {R.}~\bibnamefont
  {Kjellander}}, \bibinfo {author} {\bibfnamefont {S.}~\bibnamefont {Sarman}},
  \bibinfo {author} {\bibfnamefont {S.}~\bibnamefont {Chodankar}}, \bibinfo
  {author} {\bibfnamefont {E.}~\bibnamefont {Perret}}, \bibinfo {author}
  {\bibfnamefont {J.}~\bibnamefont {Buitenhuis}},\ and\ \bibinfo {author}
  {\bibfnamefont {J.~F.}\ \bibnamefont {van~der Veen}},\ }\bibfield  {title}
  {\enquote {\bibinfo {title} {Anisotropic pair correlations and structure
  factors of confined hard-sphere fluids: An experimental and theoretical
  study},}\ }\href {https://doi.org/10.1103/PhysRevLett.108.037802} {\bibfield
  {journal} {\bibinfo  {journal} {Phys. Rev. Lett.}\ }\textbf {\bibinfo
  {volume} {108}},\ \bibinfo {pages} {037802} (\bibinfo {year}
  {2012})}\BibitemShut {NoStop}%
\bibitem [{\citenamefont {Lang}, \citenamefont {Franosch},\ and\ \citenamefont
  {Schilling}(2014)}]{SFS14}%
  \BibitemOpen
  \bibfield  {author} {\bibinfo {author} {\bibfnamefont {S.}~\bibnamefont
  {Lang}}, \bibinfo {author} {\bibfnamefont {T.}~\bibnamefont {Franosch}},\
  and\ \bibinfo {author} {\bibfnamefont {R.}~\bibnamefont {Schilling}},\
  }\bibfield  {title} {\enquote {\bibinfo {title} {Structural quantities of
  quasi-two-dimensional fluids},}\ }\href {https://doi.org/10.1063/1.4867284}
  {\bibfield  {journal} {\bibinfo  {journal} {J. Chem. Phys.}\ }\textbf
  {\bibinfo {volume} {140}},\ \bibinfo {pages} {104506} (\bibinfo {year}
  {2014})}\BibitemShut {NoStop}%
\bibitem [{\citenamefont {Nyg{\aa}rd}\ \emph {et~al.}(2016)\citenamefont
  {Nyg{\aa}rd}, \citenamefont {Sarman}, \citenamefont {Hyltegren},
  \citenamefont {Chodankar}, \citenamefont {Perret}, \citenamefont
  {Buitenhuis}, \citenamefont {van~der Veen},\ and\ \citenamefont
  {Kjellander}}]{NSHCPBJK16}%
  \BibitemOpen
  \bibfield  {author} {\bibinfo {author} {\bibfnamefont {K.}~\bibnamefont
  {Nyg{\aa}rd}}, \bibinfo {author} {\bibfnamefont {S.}~\bibnamefont {Sarman}},
  \bibinfo {author} {\bibfnamefont {K.}~\bibnamefont {Hyltegren}}, \bibinfo
  {author} {\bibfnamefont {S.}~\bibnamefont {Chodankar}}, \bibinfo {author}
  {\bibfnamefont {E.}~\bibnamefont {Perret}}, \bibinfo {author} {\bibfnamefont
  {J.}~\bibnamefont {Buitenhuis}}, \bibinfo {author} {\bibfnamefont {J.~F.}\
  \bibnamefont {van~der Veen}},\ and\ \bibinfo {author} {\bibfnamefont
  {R.}~\bibnamefont {Kjellander}},\ }\bibfield  {title} {\enquote {\bibinfo
  {title} {Density fluctuations of hard-sphere fluids in narrow confinement},}\
  }\href {https://doi.org/10.1103/PhysRevX.6.011014} {\bibfield  {journal}
  {\bibinfo  {journal} {Phys. Rev. X}\ }\textbf {\bibinfo {volume} {6}},\
  \bibinfo {pages} {011014} (\bibinfo {year} {2016})}\BibitemShut {NoStop}%
\bibitem [{\citenamefont {Jung}\ and\ \citenamefont {Franosch}(2023)}]{JF23}%
  \BibitemOpen
  \bibfield  {author} {\bibinfo {author} {\bibfnamefont {G.}~\bibnamefont
  {Jung}}\ and\ \bibinfo {author} {\bibfnamefont {T.}~\bibnamefont
  {Franosch}},\ }\bibfield  {title} {\enquote {\bibinfo {title} {Computer
  simulations and mode-coupling theory of glass-forming confined hard-sphere
  fluids},}\ }\href {https://doi.org/10.1103/PhysRevE.107.054101} {\bibfield
  {journal} {\bibinfo  {journal} {Phys. Rev. E}\ }\textbf {\bibinfo {volume}
  {107}},\ \bibinfo {pages} {054101} (\bibinfo {year} {2023})}\BibitemShut
  {NoStop}%
\bibitem [{\citenamefont {Brey}, \citenamefont {de~Soria},\ and\ \citenamefont
  {Maynar}(2024)}]{BGM24}%
  \BibitemOpen
  \bibfield  {author} {\bibinfo {author} {\bibfnamefont {J.~J.}\ \bibnamefont
  {Brey}}, \bibinfo {author} {\bibfnamefont {M.~I.~G.}\ \bibnamefont
  {de~Soria}},\ and\ \bibinfo {author} {\bibfnamefont {P.}~\bibnamefont
  {Maynar}},\ }\bibfield  {title} {\enquote {\bibinfo {title} {Dynamics and
  kinetic theory of hard spheres under strong confinement},}\ }\href
  {https://doi.org/10.1103/PhysRevE.110.034127} {\bibfield  {journal} {\bibinfo
   {journal} {Phys. Rev. E}\ }\textbf {\bibinfo {volume} {110}},\ \bibinfo
  {pages} {034127} (\bibinfo {year} {2024})}\BibitemShut {NoStop}%
\bibitem [{\citenamefont {Huang}, \citenamefont {Yoon},\ and\ \citenamefont
  {Kwak}(2013)}]{HYS13}%
  \BibitemOpen
  \bibfield  {author} {\bibinfo {author} {\bibfnamefont {H.}~\bibnamefont
  {Huang}}, \bibinfo {author} {\bibfnamefont {Y.}~\bibnamefont {Yoon}},\ and\
  \bibinfo {author} {\bibfnamefont {S.}~\bibnamefont {Kwak}},\ }\bibfield
  {title} {\enquote {\bibinfo {title} {On the freezing and structure of hard
  spheres under spherical confinement},}\ }\href
  {https://doi.org/10.1080/00268976.2013.781694} {\bibfield  {journal}
  {\bibinfo  {journal} {Mol. Phys.}\ }\textbf {\bibinfo {volume} {111}},\
  \bibinfo {pages} {3283--3288} (\bibinfo {year} {2013})}\BibitemShut {NoStop}%
\bibitem [{\citenamefont {Wang}\ \emph {et~al.}(2021)\citenamefont {Wang},
  \citenamefont {Dasgupta}, \citenamefont {van~der Wee}, \citenamefont
  {Zanaga}, \citenamefont {Altantzis}, \citenamefont {Wu}, \citenamefont
  {Coli}, \citenamefont {Murray}, \citenamefont {Bals}, \citenamefont
  {Dijkstra},\ and\ \citenamefont {van Blaaderen}}]{WDW21}%
  \BibitemOpen
  \bibfield  {author} {\bibinfo {author} {\bibfnamefont {D.}~\bibnamefont
  {Wang}}, \bibinfo {author} {\bibfnamefont {T.}~\bibnamefont {Dasgupta}},
  \bibinfo {author} {\bibfnamefont {E.}~\bibnamefont {van~der Wee}}, \bibinfo
  {author} {\bibfnamefont {D.}~\bibnamefont {Zanaga}}, \bibinfo {author}
  {\bibfnamefont {T.}~\bibnamefont {Altantzis}}, \bibinfo {author}
  {\bibfnamefont {Y.}~\bibnamefont {Wu}}, \bibinfo {author} {\bibfnamefont
  {G.}~\bibnamefont {Coli}}, \bibinfo {author} {\bibfnamefont {C.}~\bibnamefont
  {Murray}}, \bibinfo {author} {\bibfnamefont {S.}~\bibnamefont {Bals}},
  \bibinfo {author} {\bibfnamefont {M.}~\bibnamefont {Dijkstra}},\ and\
  \bibinfo {author} {\bibfnamefont {A.}~\bibnamefont {van Blaaderen}},\
  }\bibfield  {title} {\enquote {\bibinfo {title} {Binary icosahedral clusters
  of hard spheres in spherical confinement},}\ }\href
  {https://doi.org/10.1038/s41567-020-1003-9} {\bibfield  {journal} {\bibinfo
  {journal} {Nat. Phys.}\ }\textbf {\bibinfo {volume} {17}},\ \bibinfo {pages}
  {128--134} (\bibinfo {year} {2021})}\BibitemShut {NoStop}%
\bibitem [{\citenamefont {Bratko}, \citenamefont {Blum},\ and\ \citenamefont
  {Wertheim}(1989)}]{BBW89}%
  \BibitemOpen
  \bibfield  {author} {\bibinfo {author} {\bibfnamefont {D.}~\bibnamefont
  {Bratko}}, \bibinfo {author} {\bibfnamefont {L.}~\bibnamefont {Blum}},\ and\
  \bibinfo {author} {\bibfnamefont {M.~S.}\ \bibnamefont {Wertheim}},\
  }\bibfield  {title} {\enquote {\bibinfo {title} {Structure of hard sphere
  fluids in narrow cylindrical pores},}\ }\href
  {https://doi.org/10.1063/1.455922} {\bibfield  {journal} {\bibinfo  {journal}
  {J. Chem. Phys.}\ }\textbf {\bibinfo {volume} {90}},\ \bibinfo {pages}
  {2752--2757} (\bibinfo {year} {1989})}\BibitemShut {NoStop}%
\bibitem [{\citenamefont {Alejandre}, \citenamefont {Lozada-Cassou},\ and\
  \citenamefont {Degr\`eve}(1996)}]{ALD96}%
  \BibitemOpen
  \bibfield  {author} {\bibinfo {author} {\bibfnamefont {J.}~\bibnamefont
  {Alejandre}}, \bibinfo {author} {\bibfnamefont {M.}~\bibnamefont
  {Lozada-Cassou}},\ and\ \bibinfo {author} {\bibfnamefont {L.}~\bibnamefont
  {Degr\`eve}},\ }\bibfield  {title} {\enquote {\bibinfo {title} {Effect of
  pore geometry on a confined hard sphere fluid},}\ }\href
  {https://doi.org/10.1080/00268979609484513} {\bibfield  {journal} {\bibinfo
  {journal} {Mol. Phys.}\ }\textbf {\bibinfo {volume} {88}},\ \bibinfo {pages}
  {1317--1336} (\bibinfo {year} {1996})}\BibitemShut {NoStop}%
\bibitem [{\citenamefont {Duda}\ \emph {et~al.}(1998)\citenamefont {Duda},
  \citenamefont {Sokolowski}, \citenamefont {Bryk},\ and\ \citenamefont
  {Pizio}}]{DSBP98}%
  \BibitemOpen
  \bibfield  {author} {\bibinfo {author} {\bibfnamefont {Y.}~\bibnamefont
  {Duda}}, \bibinfo {author} {\bibfnamefont {S.}~\bibnamefont {Sokolowski}},
  \bibinfo {author} {\bibfnamefont {P.}~\bibnamefont {Bryk}},\ and\ \bibinfo
  {author} {\bibfnamefont {O.}~\bibnamefont {Pizio}},\ }\bibfield  {title}
  {\enquote {\bibinfo {title} {Structure and adsorption of a hard sphere fluid
  in a cylindrical and spherical pore filled by a disordered matrix: {A}
  {M}onte {C}arlo study},}\ }\href {https://doi.org/10.1021/jp9811272}
  {\bibfield  {journal} {\bibinfo  {journal} {J. Phys. Chem. B}\ }\textbf
  {\bibinfo {volume} {102}},\ \bibinfo {pages} {5490--5494} (\bibinfo {year}
  {1998})}\BibitemShut {NoStop}%
\bibitem [{\citenamefont {Malescio}\ and\ \citenamefont
  {Pellicane}(2001)}]{MP01}%
  \BibitemOpen
  \bibfield  {author} {\bibinfo {author} {\bibfnamefont {G.}~\bibnamefont
  {Malescio}}\ and\ \bibinfo {author} {\bibfnamefont {G.}~\bibnamefont
  {Pellicane}},\ }\bibfield  {title} {\enquote {\bibinfo {title} {Simple fluids
  with complex phase behavior},}\ }\href
  {https://doi.org/10.1103/PhysRevE.63.020501} {\bibfield  {journal} {\bibinfo
  {journal} {Phys. Rev. E}\ }\textbf {\bibinfo {volume} {63}},\ \bibinfo
  {pages} {020501(R)} (\bibinfo {year} {2001})}\BibitemShut {NoStop}%
\bibitem [{\citenamefont {Kamalvand}, \citenamefont {Keshavarzi},\ and\
  \citenamefont {Mansoori}(2008)}]{KKM08}%
  \BibitemOpen
  \bibfield  {author} {\bibinfo {author} {\bibfnamefont {M.}~\bibnamefont
  {Kamalvand}}, \bibinfo {author} {\bibfnamefont {T.}~\bibnamefont
  {Keshavarzi}},\ and\ \bibinfo {author} {\bibfnamefont {G.~A.}\ \bibnamefont
  {Mansoori}},\ }\bibfield  {title} {\enquote {\bibinfo {title} {Behavior of
  the confined hard-sphere fluid within nanoslits: {A} fundamental-measure
  density-functional theory study},}\ }\href
  {https://doi.org/10.1142/S0219581X08005365} {\bibfield  {journal} {\bibinfo
  {journal} {Int. J. Nanosci.}\ }\textbf {\bibinfo {volume} {07}},\ \bibinfo
  {pages} {245--253} (\bibinfo {year} {2008})}\BibitemShut {NoStop}%
\bibitem [{\citenamefont {Dur\'an-Olivencia}\ and\ \citenamefont
  {Gordillo}(2009)}]{DG09}%
  \BibitemOpen
  \bibfield  {author} {\bibinfo {author} {\bibfnamefont {F.~J.}\ \bibnamefont
  {Dur\'an-Olivencia}}\ and\ \bibinfo {author} {\bibfnamefont {M.~C.}\
  \bibnamefont {Gordillo}},\ }\bibfield  {title} {\enquote {\bibinfo {title}
  {Ordering of hard spheres inside hard cylindrical pores},}\ }\href
  {https://doi.org/10.1103/PhysRevE.79.061111} {\bibfield  {journal} {\bibinfo
  {journal} {Phys. Rev. E}\ }\textbf {\bibinfo {volume} {79}},\ \bibinfo
  {pages} {061111} (\bibinfo {year} {2009})}\BibitemShut {NoStop}%
\bibitem [{\citenamefont {Huang}, \citenamefont {Kwak},\ and\ \citenamefont
  {Singh}(2009)}]{HKS09}%
  \BibitemOpen
  \bibfield  {author} {\bibinfo {author} {\bibfnamefont {H.~C.}\ \bibnamefont
  {Huang}}, \bibinfo {author} {\bibfnamefont {S.~K.}\ \bibnamefont {Kwak}},\
  and\ \bibinfo {author} {\bibfnamefont {J.~K.}\ \bibnamefont {Singh}},\
  }\bibfield  {title} {\enquote {\bibinfo {title} {Characterization of
  fluid-solid phase transition of hard-sphere fluids in cylindrical pore via
  molecular dynamics simulation},}\ }\href {https://doi.org/10.1063/1.3120486}
  {\bibfield  {journal} {\bibinfo  {journal} {J. Chem. Phys.}\ }\textbf
  {\bibinfo {volume} {130}},\ \bibinfo {pages} {164511} (\bibinfo {year}
  {2009})}\BibitemShut {NoStop}%
\bibitem [{\citenamefont {Mandal}\ \emph {et~al.}(2014)\citenamefont {Mandal},
  \citenamefont {Lang}, \citenamefont {Gross}, \citenamefont {Oettel},
  \citenamefont {Raabe}, \citenamefont {Franosch},\ and\ \citenamefont
  {Varnik}}]{MLGORFV14}%
  \BibitemOpen
  \bibfield  {author} {\bibinfo {author} {\bibfnamefont {S.}~\bibnamefont
  {Mandal}}, \bibinfo {author} {\bibfnamefont {S.}~\bibnamefont {Lang}},
  \bibinfo {author} {\bibfnamefont {M.}~\bibnamefont {Gross}}, \bibinfo
  {author} {\bibfnamefont {M.}~\bibnamefont {Oettel}}, \bibinfo {author}
  {\bibfnamefont {D.}~\bibnamefont {Raabe}}, \bibinfo {author} {\bibfnamefont
  {T.}~\bibnamefont {Franosch}},\ and\ \bibinfo {author} {\bibfnamefont
  {F.}~\bibnamefont {Varnik}},\ }\bibfield  {title} {\enquote {\bibinfo {title}
  {Multiple reentrant glass transitions in confined hard-sphere glasses},}\
  }\href {https://doi.org/doi.org/10.1038/ncomms5435} {\bibfield  {journal}
  {\bibinfo  {journal} {Nat. Commun.}\ }\textbf {\bibinfo {volume} {5}},\
  \bibinfo {pages} {4435} (\bibinfo {year} {2014})}\BibitemShut {NoStop}%
\bibitem [{\citenamefont {Godfrey}\ and\ \citenamefont {Moore}(2014)}]{GM14}%
  \BibitemOpen
  \bibfield  {author} {\bibinfo {author} {\bibfnamefont {M.~J.}\ \bibnamefont
  {Godfrey}}\ and\ \bibinfo {author} {\bibfnamefont {M.~A.}\ \bibnamefont
  {Moore}},\ }\bibfield  {title} {\enquote {\bibinfo {title} {Static and
  dynamical properties of a hard-disk fluid confined to a narrow channel},}\
  }\href {https://doi.org/10.1103/PhysRevE.89.032111} {\bibfield  {journal}
  {\bibinfo  {journal} {Phys. Rev. E}\ }\textbf {\bibinfo {volume} {89}},\
  \bibinfo {pages} {032111} (\bibinfo {year} {2014})}\BibitemShut {NoStop}%
\bibitem [{\citenamefont {Krapivsky}, \citenamefont {Mallick},\ and\
  \citenamefont {Sadhu}(2014)}]{KMS14}%
  \BibitemOpen
  \bibfield  {author} {\bibinfo {author} {\bibfnamefont {P.~L.}\ \bibnamefont
  {Krapivsky}}, \bibinfo {author} {\bibfnamefont {K.}~\bibnamefont {Mallick}},\
  and\ \bibinfo {author} {\bibfnamefont {T.}~\bibnamefont {Sadhu}},\ }\bibfield
   {title} {\enquote {\bibinfo {title} {Large deviations in single-file
  diffusion},}\ }\href {https://doi.org/10.1103/PhysRevLett.113.078101}
  {\bibfield  {journal} {\bibinfo  {journal} {Phys. Rev. Lett.}\ }\textbf
  {\bibinfo {volume} {113}},\ \bibinfo {pages} {078101} (\bibinfo {year}
  {2014})}\BibitemShut {NoStop}%
\bibitem [{\citenamefont {Wittmann}, \citenamefont {L\"owen},\ and\
  \citenamefont {Brader}(2021)}]{WLB20}%
  \BibitemOpen
  \bibfield  {author} {\bibinfo {author} {\bibfnamefont {R.}~\bibnamefont
  {Wittmann}}, \bibinfo {author} {\bibfnamefont {H.}~\bibnamefont {L\"owen}},\
  and\ \bibinfo {author} {\bibfnamefont {J.~M.}\ \bibnamefont {Brader}},\
  }\bibfield  {title} {\enquote {\bibinfo {title} {Order-preserving dynamics in
  one dimension -- single-file diffusion and caging from the perspective of
  dynamical density functional theory},}\ }\href
  {https://doi.org/10.1080/00268976.2020.1867250} {\bibfield  {journal}
  {\bibinfo  {journal} {Mol. Phys.}\ }\textbf {\bibinfo {volume} {119}},\
  \bibinfo {pages} {e1867250} (\bibinfo {year} {2021})}\BibitemShut {NoStop}%
\bibitem [{\citenamefont {Maynar}, \citenamefont {{Garc\'ia de Soria}},\ and\
  \citenamefont {Brey}(2022)}]{MGB22}%
  \BibitemOpen
  \bibfield  {author} {\bibinfo {author} {\bibfnamefont {P.}~\bibnamefont
  {Maynar}}, \bibinfo {author} {\bibfnamefont {M.~I.}\ \bibnamefont {{Garc\'ia
  de Soria}}},\ and\ \bibinfo {author} {\bibfnamefont {J.~J.}\ \bibnamefont
  {Brey}},\ }\bibfield  {title} {\enquote {\bibinfo {title} {Dynamics of an
  inelastic tagged particle under strong confinement},}\ }\href
  {https://doi.org/10.1063/5.0129279} {\bibfield  {journal} {\bibinfo
  {journal} {Phys. Fluids}\ }\textbf {\bibinfo {volume} {34}},\ \bibinfo
  {pages} {123321} (\bibinfo {year} {2022})}\BibitemShut {NoStop}%
\bibitem [{\citenamefont {Salsburg}, \citenamefont {Zwanzig},\ and\
  \citenamefont {Kirkwood}(1953)}]{SZK53}%
  \BibitemOpen
  \bibfield  {author} {\bibinfo {author} {\bibfnamefont {Z.~W.}\ \bibnamefont
  {Salsburg}}, \bibinfo {author} {\bibfnamefont {R.~W.}\ \bibnamefont
  {Zwanzig}},\ and\ \bibinfo {author} {\bibfnamefont {J.~G.}\ \bibnamefont
  {Kirkwood}},\ }\bibfield  {title} {\enquote {\bibinfo {title} {Molecular
  distribution functions in a one-dimensional fluid},}\ }\href
  {https://doi.org/10.1063/1.1699116} {\bibfield  {journal} {\bibinfo
  {journal} {J. Chem. Phys.}\ }\textbf {\bibinfo {volume} {21}},\ \bibinfo
  {pages} {1098--1107} (\bibinfo {year} {1953})}\BibitemShut {NoStop}%
\bibitem [{\citenamefont {Kikuchi}(1955)}]{K55b}%
  \BibitemOpen
  \bibfield  {author} {\bibinfo {author} {\bibfnamefont {R.}~\bibnamefont
  {Kikuchi}},\ }\bibfield  {title} {\enquote {\bibinfo {title} {Theory of
  one-dimensional fluid binary mixtures},}\ }\href
  {https://doi.org/10.1063/1.1741874} {\bibfield  {journal} {\bibinfo
  {journal} {J. Chem. Phys.}\ }\textbf {\bibinfo {volume} {23}},\ \bibinfo
  {pages} {2327--2332} (\bibinfo {year} {1955})}\BibitemShut {NoStop}%
\bibitem [{\citenamefont {Katsura}\ and\ \citenamefont {Tago}(1968)}]{KT68}%
  \BibitemOpen
  \bibfield  {author} {\bibinfo {author} {\bibfnamefont {S.}~\bibnamefont
  {Katsura}}\ and\ \bibinfo {author} {\bibfnamefont {Y.}~\bibnamefont {Tago}},\
  }\bibfield  {title} {\enquote {\bibinfo {title} {Radial distribution function
  and the direct correlation function for one-dimensional gas with square-well
  potential},}\ }\href {https://doi.org/10.1063/1.1669764} {\bibfield
  {journal} {\bibinfo  {journal} {J. Chem. Phys.}\ }\textbf {\bibinfo {volume}
  {48}},\ \bibinfo {pages} {4246--4251} (\bibinfo {year} {1968})}\BibitemShut
  {NoStop}%
\bibitem [{\citenamefont {Rybicki}(1971)}]{R71b}%
  \BibitemOpen
  \bibfield  {author} {\bibinfo {author} {\bibfnamefont {G.~B.}\ \bibnamefont
  {Rybicki}},\ }\bibfield  {title} {\enquote {\bibinfo {title} {Exact
  statistical mechanics of a one-dimensional self-gravitating system},}\ }\href
  {https://doi.org/10.1007/BF00649195} {\bibfield  {journal} {\bibinfo
  {journal} {Astrophys. Space Sci.}\ }\textbf {\bibinfo {volume} {14}},\
  \bibinfo {pages} {56--72} (\bibinfo {year} {1971})}\BibitemShut {NoStop}%
\bibitem [{\citenamefont {Percus}(1982)}]{P82}%
  \BibitemOpen
  \bibfield  {author} {\bibinfo {author} {\bibfnamefont {J.~K.}\ \bibnamefont
  {Percus}},\ }\bibfield  {title} {\enquote {\bibinfo {title} {One-dimensional
  classical fluid with nearest-neighbor interaction in arbitrary external
  field},}\ }\href {https://doi.org/10.1007/BF01011623} {\bibfield  {journal}
  {\bibinfo  {journal} {J. Stat. Phys.}\ }\textbf {\bibinfo {volume} {28}},\
  \bibinfo {pages} {67--81} (\bibinfo {year} {1982})}\BibitemShut {NoStop}%
\bibitem [{\citenamefont {Bishop}\ and\ \citenamefont
  {Boonstra}(1983)}]{BB83a}%
  \BibitemOpen
  \bibfield  {author} {\bibinfo {author} {\bibfnamefont {M.}~\bibnamefont
  {Bishop}}\ and\ \bibinfo {author} {\bibfnamefont {M.~A.}\ \bibnamefont
  {Boonstra}},\ }\bibfield  {title} {\enquote {\bibinfo {title} {Exact
  partition functions for some one-dimensional models via the isobaric
  ensemble},}\ }\href {https://doi.org/10.1119/1.13204} {\bibfield  {journal}
  {\bibinfo  {journal} {Am. J. Phys.}\ }\textbf {\bibinfo {volume} {51}},\
  \bibinfo {pages} {564--566} (\bibinfo {year} {1983})}\BibitemShut {NoStop}%
\bibitem [{\citenamefont {Heying}\ and\ \citenamefont {Corti}(2004)}]{HC04}%
  \BibitemOpen
  \bibfield  {author} {\bibinfo {author} {\bibfnamefont {M.}~\bibnamefont
  {Heying}}\ and\ \bibinfo {author} {\bibfnamefont {D.~S.}\ \bibnamefont
  {Corti}},\ }\bibfield  {title} {\enquote {\bibinfo {title} {The
  one-dimensional fully non-additive binary hard rod mixture: exact
  thermophysical properties},}\ }\href
  {https://doi.org/10.1016/j.fluid.2004.02.018} {\bibfield  {journal} {\bibinfo
   {journal} {Fluid Phase Equilib.}\ }\textbf {\bibinfo {volume} {220}},\
  \bibinfo {pages} {85--103} (\bibinfo {year} {2004})}\BibitemShut {NoStop}%
\bibitem [{\citenamefont {Santos}(2007)}]{S07}%
  \BibitemOpen
  \bibfield  {author} {\bibinfo {author} {\bibfnamefont {A.}~\bibnamefont
  {Santos}},\ }\bibfield  {title} {\enquote {\bibinfo {title} {Exact bulk
  correlation functions in one-dimensional nonadditive hard-core mixtures},}\
  }\href {https://doi.org/10.1103/PhysRevE.76.062201} {\bibfield  {journal}
  {\bibinfo  {journal} {Phys. Rev. E}\ }\textbf {\bibinfo {volume} {76}},\
  \bibinfo {pages} {{062}{201}} (\bibinfo {year} {2007})}\BibitemShut {NoStop}%
\bibitem [{\citenamefont {Ben-Naim}\ and\ \citenamefont
  {Santos}(2009)}]{BNS09}%
  \BibitemOpen
  \bibfield  {author} {\bibinfo {author} {\bibfnamefont {A.}~\bibnamefont
  {Ben-Naim}}\ and\ \bibinfo {author} {\bibfnamefont {A.}~\bibnamefont
  {Santos}},\ }\bibfield  {title} {\enquote {\bibinfo {title} {Local and global
  properties of mixtures in one-dimensional systems. {II}. {Exact} results for
  the {Kirkwood}--{Buff} integrals},}\ }\href
  {https://doi.org/10.1063/1.3256234} {\bibfield  {journal} {\bibinfo
  {journal} {J. Chem. Phys.}\ }\textbf {\bibinfo {volume} {131}},\ \bibinfo
  {pages} {{164}{512}} (\bibinfo {year} {2009})}\BibitemShut {NoStop}%
\bibitem [{\citenamefont {Fantoni}(2016)}]{F16}%
  \BibitemOpen
  \bibfield  {author} {\bibinfo {author} {\bibfnamefont {R.}~\bibnamefont
  {Fantoni}},\ }\bibfield  {title} {\enquote {\bibinfo {title} {Exact results
  for one dimensional fluids through functional integration},}\ }\href
  {https://doi.org/10.1007/s10955-016-1510-3} {\bibfield  {journal} {\bibinfo
  {journal} {J. Stat. Phys.}\ }\textbf {\bibinfo {volume} {163}},\ \bibinfo
  {pages} {1247--1267} (\bibinfo {year} {2016})}\BibitemShut {NoStop}%
\bibitem [{\citenamefont {Montero}\ and\ \citenamefont {Santos}(2017)}]{MS17}%
  \BibitemOpen
  \bibfield  {author} {\bibinfo {author} {\bibfnamefont {A.~M.}\ \bibnamefont
  {Montero}}\ and\ \bibinfo {author} {\bibfnamefont {A.}~\bibnamefont
  {Santos}},\ }\href@noop {} {} (\bibinfo {year} {2017}),\ \bibinfo {note}
  {``Radial Distribution Function for One-Dimensional Triangle Well and Ramp
  Fluids'', Wolfram Demonstrations Project,
  \url{http://demonstrations.wolfram.com/RadialDistributionFunctionForOneDimensionalTriangleWellAndRa/}}\BibitemShut
  {NoStop}%
\bibitem [{\citenamefont {Montero}\ and\ \citenamefont {Santos}(2019)}]{MS19}%
  \BibitemOpen
  \bibfield  {author} {\bibinfo {author} {\bibfnamefont {A.~M.}\ \bibnamefont
  {Montero}}\ and\ \bibinfo {author} {\bibfnamefont {A.}~\bibnamefont
  {Santos}},\ }\bibfield  {title} {\enquote {\bibinfo {title} {Triangle-well
  and ramp interactions in one-dimensional fluids: A fully analytic exact
  solution},}\ }\href {https://doi.org/10.1007/s10955-019-02255-x} {\bibfield
  {journal} {\bibinfo  {journal} {J. Stat. Phys.}\ }\textbf {\bibinfo {volume}
  {175}},\ \bibinfo {pages} {269--288} (\bibinfo {year} {2019})}\BibitemShut
  {NoStop}%
\bibitem [{\citenamefont {Maestre}\ and\ \citenamefont {Santos}(2020)}]{MS20}%
  \BibitemOpen
  \bibfield  {author} {\bibinfo {author} {\bibfnamefont {M.~A.~G.}\
  \bibnamefont {Maestre}}\ and\ \bibinfo {author} {\bibfnamefont
  {A.}~\bibnamefont {Santos}},\ }\bibfield  {title} {\enquote {\bibinfo {title}
  {One-dimensional {J}anus fluids. {E}xact solution and mapping from the
  quenched to the annealed system},}\ }\href
  {https://doi.org/10.1088/1742-5468/ab900d} {\bibfield  {journal} {\bibinfo
  {journal} {J. Stat. Mech.}\ }\textbf {\bibinfo {volume} {2020}},\ \bibinfo
  {pages} {063217} (\bibinfo {year} {2020})}\BibitemShut {NoStop}%
\bibitem [{\citenamefont {Barker}(1962)}]{B62}%
  \BibitemOpen
  \bibfield  {author} {\bibinfo {author} {\bibfnamefont {J.~A.}\ \bibnamefont
  {Barker}},\ }\bibfield  {title} {\enquote {\bibinfo {title} {Statistical
  mechanics of almost one-dimensional systems},}\ }\href
  {https://doi.org/10.1071/PH620127} {\bibfield  {journal} {\bibinfo  {journal}
  {Aust. J. Phys.,}\ }\textbf {\bibinfo {volume} {15}},\ \bibinfo {pages}
  {127--134} (\bibinfo {year} {1962})}\BibitemShut {NoStop}%
\bibitem [{\citenamefont {Barker}(1964)}]{B64b}%
  \BibitemOpen
  \bibfield  {author} {\bibinfo {author} {\bibfnamefont {J.}~\bibnamefont
  {Barker}},\ }\bibfield  {title} {\enquote {\bibinfo {title} {Statistical
  mechanics of almost one-dimensional systems. {II}},}\ }\href
  {https://doi.org/10.1071/PH640259} {\bibfield  {journal} {\bibinfo  {journal}
  {Aust. J. Phys.,}\ }\textbf {\bibinfo {volume} {17}},\ \bibinfo {pages}
  {259--268} (\bibinfo {year} {1964})}\BibitemShut {NoStop}%
\bibitem [{\citenamefont {Kofke}\ and\ \citenamefont {Post}(1993)}]{KP93}%
  \BibitemOpen
  \bibfield  {author} {\bibinfo {author} {\bibfnamefont {D.~A.}\ \bibnamefont
  {Kofke}}\ and\ \bibinfo {author} {\bibfnamefont {A.~J.}\ \bibnamefont
  {Post}},\ }\bibfield  {title} {\enquote {\bibinfo {title} {Hard particles in
  narrow pores. {T}ransfer-matrix solution and the periodic narrow box},}\
  }\href {https://doi.org/10.1063/1.464967} {\bibfield  {journal} {\bibinfo
  {journal} {J. Chem. Phys.}\ }\textbf {\bibinfo {volume} {98}},\ \bibinfo
  {pages} {4853--4861} (\bibinfo {year} {1993})}\BibitemShut {NoStop}%
\bibitem [{\citenamefont {K\"ofinger}, \citenamefont {Hummer},\ and\
  \citenamefont {Dellago}(2011)}]{KHD11}%
  \BibitemOpen
  \bibfield  {author} {\bibinfo {author} {\bibfnamefont {J.}~\bibnamefont
  {K\"ofinger}}, \bibinfo {author} {\bibfnamefont {G.}~\bibnamefont {Hummer}},\
  and\ \bibinfo {author} {\bibfnamefont {C.}~\bibnamefont {Dellago}},\
  }\bibfield  {title} {\enquote {\bibinfo {title} {Single-file water in
  nanopores},}\ }\href {https://doi.org/10.1039/C1CP21086F} {\bibfield
  {journal} {\bibinfo  {journal} {Phys. Chem. Chem. Phys.}\ }\textbf {\bibinfo
  {volume} {13}},\ \bibinfo {pages} {15403--15417} (\bibinfo {year}
  {2011})}\BibitemShut {NoStop}%
\bibitem [{\citenamefont {Manning}(2024)}]{M24}%
  \BibitemOpen
  \bibfield  {author} {\bibinfo {author} {\bibfnamefont {G.~S.}\ \bibnamefont
  {Manning}},\ }\bibfield  {title} {\enquote {\bibinfo {title} {A hard sphere
  model for single-file water transport across biological membranes},}\ }\href
  {https://doi.org/10.1140/epje/s10189-024-00419-6} {\bibfield  {journal}
  {\bibinfo  {journal} {Eur. Phys. J. E}\ }\textbf {\bibinfo {volume} {47}},\
  \bibinfo {pages} {27} (\bibinfo {year} {2024})}\BibitemShut {NoStop}%
\bibitem [{\citenamefont {Kamenetskiy}, \citenamefont {Mon},\ and\
  \citenamefont {Percus}(2004)}]{KMP04}%
  \BibitemOpen
  \bibfield  {author} {\bibinfo {author} {\bibfnamefont {I.~E.}\ \bibnamefont
  {Kamenetskiy}}, \bibinfo {author} {\bibfnamefont {K.~K.}\ \bibnamefont
  {Mon}},\ and\ \bibinfo {author} {\bibfnamefont {J.~K.}\ \bibnamefont
  {Percus}},\ }\bibfield  {title} {\enquote {\bibinfo {title} {Equation of
  state for hard-sphere fluid in restricted geometry},}\ }\href
  {https://doi.org/10.1063/1.1795131} {\bibfield  {journal} {\bibinfo
  {journal} {J. Chem. Phys.}\ }\textbf {\bibinfo {volume} {121}},\ \bibinfo
  {pages} {7355--7361} (\bibinfo {year} {2004})}\BibitemShut {NoStop}%
\bibitem [{\citenamefont {Gurin}\ and\ \citenamefont {Varga}(2013)}]{GV13}%
  \BibitemOpen
  \bibfield  {author} {\bibinfo {author} {\bibfnamefont {P.}~\bibnamefont
  {Gurin}}\ and\ \bibinfo {author} {\bibfnamefont {S.}~\bibnamefont {Varga}},\
  }\bibfield  {title} {\enquote {\bibinfo {title} {Pair correlation functions
  of two- and three-dimensional hard-core fluids confined into narrow pores:
  {E}xact results from transfer-matrix method},}\ }\href
  {https://doi.org/10.1063/1.4852181} {\bibfield  {journal} {\bibinfo
  {journal} {J. Chem. Phys.}\ }\textbf {\bibinfo {volume} {139}},\ \bibinfo
  {pages} {244708} (\bibinfo {year} {2013})}\BibitemShut {NoStop}%
\bibitem [{\citenamefont {Hu}, \citenamefont {Fu},\ and\ \citenamefont
  {Charbonneau}(2018)}]{HFC18}%
  \BibitemOpen
  \bibfield  {author} {\bibinfo {author} {\bibfnamefont {Y.}~\bibnamefont
  {Hu}}, \bibinfo {author} {\bibfnamefont {L.}~\bibnamefont {Fu}},\ and\
  \bibinfo {author} {\bibfnamefont {P.}~\bibnamefont {Charbonneau}},\
  }\bibfield  {title} {\enquote {\bibinfo {title} {Correlation lengths in
  quasi-one-dimensional systems via transfer matrices},}\ }\href
  {https://doi.org/10.1080/00268976.2018.1479543} {\bibfield  {journal}
  {\bibinfo  {journal} {Mol. Phys.}\ }\textbf {\bibinfo {volume} {116}},\
  \bibinfo {pages} {3345--3354} (\bibinfo {year} {2018})}\BibitemShut {NoStop}%
\bibitem [{\citenamefont {Mon}(2018)}]{M18}%
  \BibitemOpen
  \bibfield  {author} {\bibinfo {author} {\bibfnamefont {K.~K.}\ \bibnamefont
  {Mon}},\ }\bibfield  {title} {\enquote {\bibinfo {title} {Virial series
  expansion and {M}onte {C}arlo studies of equation of state for hard spheres
  in narrow cylindrical pores},}\ }\href
  {https://doi.org/10.1103/PhysRevE.97.052114} {\bibfield  {journal} {\bibinfo
  {journal} {Phys. Rev. E}\ }\textbf {\bibinfo {volume} {97}},\ \bibinfo
  {pages} {052114} (\bibinfo {year} {2018})}\BibitemShut {NoStop}%
\bibitem [{\citenamefont {Nikolaev}(2019)}]{N19}%
  \BibitemOpen
  \bibfield  {author} {\bibinfo {author} {\bibfnamefont {P.~N.}\ \bibnamefont
  {Nikolaev}},\ }\bibfield  {title} {\enquote {\bibinfo {title} {Free energy
  and the equation of state of a system of solid spheres in narrow cylindrical
  pores},}\ }\href {https://doi.org/10.3103/S0027134919020140} {\bibfield
  {journal} {\bibinfo  {journal} {Moscow Univ. Phys. Bull.}\ }\textbf {\bibinfo
  {volume} {74}},\ \bibinfo {pages} {124--130} (\bibinfo {year}
  {2019})}\BibitemShut {NoStop}%
\bibitem [{\citenamefont {Franosch}\ and\ \citenamefont
  {Schilling}(2024)}]{FS24}%
  \BibitemOpen
  \bibfield  {author} {\bibinfo {author} {\bibfnamefont {T.}~\bibnamefont
  {Franosch}}\ and\ \bibinfo {author} {\bibfnamefont {R.}~\bibnamefont
  {Schilling}},\ }\bibfield  {title} {\enquote {\bibinfo {title} {Thermodynamic
  properties of quasi-one-dimensional fluids},}\ }\href
  {https://doi.org/224504} {\bibfield  {journal} {\bibinfo  {journal} {J. Chem.
  Phys.}\ }\textbf {\bibinfo {volume} {160}},\ \bibinfo {pages} {224504}
  (\bibinfo {year} {2024})}\BibitemShut {NoStop}%
\bibitem [{\citenamefont {Pickett}, \citenamefont {Gross},\ and\ \citenamefont
  {Okuyama}(2000)}]{PGO00}%
  \BibitemOpen
  \bibfield  {author} {\bibinfo {author} {\bibfnamefont {G.~T.}\ \bibnamefont
  {Pickett}}, \bibinfo {author} {\bibfnamefont {M.}~\bibnamefont {Gross}},\
  and\ \bibinfo {author} {\bibfnamefont {H.}~\bibnamefont {Okuyama}},\
  }\bibfield  {title} {\enquote {\bibinfo {title} {Spontaneous chirality in
  simple systems},}\ }\href {https://doi.org/10.1103/PhysRevLett.85.3652}
  {\bibfield  {journal} {\bibinfo  {journal} {Phys. Rev. Lett.}\ }\textbf
  {\bibinfo {volume} {85}},\ \bibinfo {pages} {3652--3655} (\bibinfo {year}
  {2000})}\BibitemShut {NoStop}%
\bibitem [{\citenamefont {Chan}(2011)}]{C11}%
  \BibitemOpen
  \bibfield  {author} {\bibinfo {author} {\bibfnamefont {H.-K.}\ \bibnamefont
  {Chan}},\ }\bibfield  {title} {\enquote {\bibinfo {title} {Densest columnar
  structures of hard spheres from sequential deposition},}\ }\href
  {https://doi.org/10.1103/PhysRevE.84.050302} {\bibfield  {journal} {\bibinfo
  {journal} {Phys. Rev. E}\ }\textbf {\bibinfo {volume} {84}},\ \bibinfo
  {pages} {050302} (\bibinfo {year} {2011})}\BibitemShut {NoStop}%
\bibitem [{\citenamefont {Mughal}, \citenamefont {Chan},\ and\ \citenamefont
  {Weaire}(2011)}]{MCW11}%
  \BibitemOpen
  \bibfield  {author} {\bibinfo {author} {\bibfnamefont {A.}~\bibnamefont
  {Mughal}}, \bibinfo {author} {\bibfnamefont {H.-K.}\ \bibnamefont {Chan}},\
  and\ \bibinfo {author} {\bibfnamefont {D.}~\bibnamefont {Weaire}},\
  }\bibfield  {title} {\enquote {\bibinfo {title} {Phyllotactic description of
  hard sphere packing in cylindrical channels},}\ }\href
  {https://doi.org/10.1103/PhysRevLett.106.115704} {\bibfield  {journal}
  {\bibinfo  {journal} {Phys. Rev. Lett.}\ }\textbf {\bibinfo {volume} {106}},\
  \bibinfo {pages} {115704} (\bibinfo {year} {2011})}\BibitemShut {NoStop}%
\bibitem [{\citenamefont {Mughal}\ \emph {et~al.}(2012)\citenamefont {Mughal},
  \citenamefont {Chan}, \citenamefont {Weaire},\ and\ \citenamefont
  {Hutzler}}]{MCWH12}%
  \BibitemOpen
  \bibfield  {author} {\bibinfo {author} {\bibfnamefont {A.}~\bibnamefont
  {Mughal}}, \bibinfo {author} {\bibfnamefont {H.-K.}\ \bibnamefont {Chan}},
  \bibinfo {author} {\bibfnamefont {D.}~\bibnamefont {Weaire}},\ and\ \bibinfo
  {author} {\bibfnamefont {S.}~\bibnamefont {Hutzler}},\ }\bibfield  {title}
  {\enquote {\bibinfo {title} {Dense packings of spheres in cylinders:
  {S}imulations},}\ }\href {https://doi.org/10.1103/PhysRevE.85.051305}
  {\bibfield  {journal} {\bibinfo  {journal} {Phys. Rev. E}\ }\textbf {\bibinfo
  {volume} {85}},\ \bibinfo {pages} {051305} (\bibinfo {year}
  {2012})}\BibitemShut {NoStop}%
\bibitem [{\citenamefont {Yamchi}\ and\ \citenamefont {Bowles}(2015)}]{YB15}%
  \BibitemOpen
  \bibfield  {author} {\bibinfo {author} {\bibfnamefont {M.~Z.}\ \bibnamefont
  {Yamchi}}\ and\ \bibinfo {author} {\bibfnamefont {R.~K.}\ \bibnamefont
  {Bowles}},\ }\bibfield  {title} {\enquote {\bibinfo {title} {Helical defect
  packings in a quasi-one-dimensional system of cylindrically confined hard
  spheres},}\ }\href {https://doi.org/10.1103/PhysRevLett.115.025702}
  {\bibfield  {journal} {\bibinfo  {journal} {Phys. Rev. Lett.}\ }\textbf
  {\bibinfo {volume} {115}},\ \bibinfo {pages} {025702} (\bibinfo {year}
  {2015})}\BibitemShut {NoStop}%
\bibitem [{\citenamefont {Fu}\ \emph {et~al.}(2016)\citenamefont {Fu},
  \citenamefont {W.~Steinhardt}, \citenamefont {Socolar},\ and\ \citenamefont
  {Charbonneau}}]{FSZSC16}%
  \BibitemOpen
  \bibfield  {author} {\bibinfo {author} {\bibfnamefont {L.}~\bibnamefont
  {Fu}}, \bibinfo {author} {\bibfnamefont {a.~H.~Z.}\ \bibnamefont
  {W.~Steinhardt}}, \bibinfo {author} {\bibfnamefont {J.~E.~S.}\ \bibnamefont
  {Socolar}},\ and\ \bibinfo {author} {\bibfnamefont {P.}~\bibnamefont
  {Charbonneau}},\ }\bibfield  {title} {\enquote {\bibinfo {title} {Hard sphere
  packings within cylinders},}\ }\href {https://doi.org/10.1039/C5SM02875B}
  {\bibfield  {journal} {\bibinfo  {journal} {Soft Matter}\ }\textbf {\bibinfo
  {volume} {12}},\ \bibinfo {pages} {2505--2514} (\bibinfo {year}
  {2016})}\BibitemShut {NoStop}%
\bibitem [{\citenamefont {Fu}\ \emph {et~al.}(2017)\citenamefont {Fu},
  \citenamefont {Bian}, \citenamefont {Shields}, \citenamefont {Cruz},
  \citenamefont {L\'opez},\ and\ \citenamefont {Charbonneau}}]{FBSCLC17}%
  \BibitemOpen
  \bibfield  {author} {\bibinfo {author} {\bibfnamefont {L.}~\bibnamefont
  {Fu}}, \bibinfo {author} {\bibfnamefont {C.}~\bibnamefont {Bian}}, \bibinfo
  {author} {\bibfnamefont {C.~W.}\ \bibnamefont {Shields}}, \bibinfo {author}
  {\bibfnamefont {D.~F.}\ \bibnamefont {Cruz}}, \bibinfo {author}
  {\bibfnamefont {G.~P.}\ \bibnamefont {L\'opez}},\ and\ \bibinfo {author}
  {\bibfnamefont {P.}~\bibnamefont {Charbonneau}},\ }\bibfield  {title}
  {\enquote {\bibinfo {title} {Assembly of hard spheres in a cylinder: a
  computational and experimental study},}\ }\href
  {https://doi.org/10.1039/C7SM00316A} {\bibfield  {journal} {\bibinfo
  {journal} {Soft Matter}\ }\textbf {\bibinfo {volume} {13}},\ \bibinfo {pages}
  {3296--3306} (\bibinfo {year} {2017})}\BibitemShut {NoStop}%
\bibitem [{\citenamefont {Ma}\ and\ \citenamefont {Chan}(2021)}]{MC21}%
  \BibitemOpen
  \bibfield  {author} {\bibinfo {author} {\bibfnamefont {P.}~\bibnamefont
  {Ma}}\ and\ \bibinfo {author} {\bibfnamefont {H.-K.}\ \bibnamefont {Chan}},\
  }\bibfield  {title} {\enquote {\bibinfo {title} {Densest-packed columnar
  structures of hard spheres: {A}n investigation of the structural dependence
  of electrical conductivity},}\ }\href
  {https://doi.org/10.3389/fphy.2021.778001} {\bibfield  {journal} {\bibinfo
  {journal} {Front. Phys.}\ }\textbf {\bibinfo {volume} {9}},\ \bibinfo {pages}
  {778001} (\bibinfo {year} {2021})}\BibitemShut {NoStop}%
\bibitem [{\citenamefont {Zarif}, \citenamefont {Spiteri},\ and\ \citenamefont
  {Bowles}(2021)}]{ZSB21}%
  \BibitemOpen
  \bibfield  {author} {\bibinfo {author} {\bibfnamefont {M.}~\bibnamefont
  {Zarif}}, \bibinfo {author} {\bibfnamefont {R.~J.}\ \bibnamefont {Spiteri}},\
  and\ \bibinfo {author} {\bibfnamefont {R.~K.}\ \bibnamefont {Bowles}},\
  }\bibfield  {title} {\enquote {\bibinfo {title} {Inherent structure landscape
  of hard spheres confined to narrow cylindrical channels},}\ }\href
  {https://doi.org/10.1103/PhysRevE.104.064602} {\bibfield  {journal} {\bibinfo
   {journal} {Phys. Rev. E}\ }\textbf {\bibinfo {volume} {104}},\ \bibinfo
  {pages} {064602} (\bibinfo {year} {2021})}\BibitemShut {NoStop}%
\bibitem [{\citenamefont {Winkelmann}\ and\ \citenamefont {Chan}(2023)}]{WC23}%
  \BibitemOpen
  \bibfield  {author} {\bibinfo {author} {\bibfnamefont {J.}~\bibnamefont
  {Winkelmann}}\ and\ \bibinfo {author} {\bibfnamefont {H.-K.}\ \bibnamefont
  {Chan}},\ }\href@noop {} {\emph {\bibinfo {title} {Columnar Structures of
  Spheres: {F}undamentals and Applications}}}\ (\bibinfo  {publisher} {Jenny
  Stanford Publishing},\ \bibinfo {address} {New York},\ \bibinfo {year}
  {2023})\BibitemShut {NoStop}%
\bibitem [{\citenamefont {Zarif}\ and\ \citenamefont {Bowles}(0243)}]{ZB23}%
  \BibitemOpen
  \bibfield  {author} {\bibinfo {author} {\bibfnamefont {M.}~\bibnamefont
  {Zarif}}\ and\ \bibinfo {author} {\bibfnamefont {R.~K.}\ \bibnamefont
  {Bowles}},\ }\bibfield  {title} {\enquote {\bibinfo {title} {Thermodynamics,
  structure and dynamics of cylindrically confined hard spheres: {T}he role of
  excess helical twist},}\ }\href {https://doi.org/10.48550/arXiv.2306.04134}
  {\bibfield  {journal} {\bibinfo  {journal} {arXiv:2306.04134}\ } (\bibinfo
  {year} {20243}),\ 10.48550/arXiv.2306.04134}\BibitemShut {NoStop}%
\bibitem [{\citenamefont {Chan}, \citenamefont {Wang},\ and\ \citenamefont
  {Han}(2019)}]{CWH19}%
  \BibitemOpen
  \bibfield  {author} {\bibinfo {author} {\bibfnamefont {H.-K.}\ \bibnamefont
  {Chan}}, \bibinfo {author} {\bibfnamefont {Y.}~\bibnamefont {Wang}},\ and\
  \bibinfo {author} {\bibfnamefont {H.}~\bibnamefont {Han}},\ }\bibfield
  {title} {\enquote {\bibinfo {title} {Densest helical structures of hard
  spheres in narrow confinement: {A}n analytic derivation},}\ }\href
  {https://doi.org/10.1063/1.5131318} {\bibfield  {journal} {\bibinfo
  {journal} {AIP Adv.}\ }\textbf {\bibinfo {volume} {9}},\ \bibinfo {pages}
  {125118} (\bibinfo {year} {2019})}\BibitemShut {NoStop}%
\bibitem [{\citenamefont {Koga}\ and\ \citenamefont {Tanaka}(2006)}]{KT06}%
  \BibitemOpen
  \bibfield  {author} {\bibinfo {author} {\bibfnamefont {K.}~\bibnamefont
  {Koga}}\ and\ \bibinfo {author} {\bibfnamefont {H.}~\bibnamefont {Tanaka}},\
  }\bibfield  {title} {\enquote {\bibinfo {title} {Close-packed structures and
  phase diagram of soft spheres in cylindrical pores},}\ }\href
  {https://doi.org/10.1063/1.2172592} {\bibfield  {journal} {\bibinfo
  {journal} {J. Chem. Phys.}\ }\textbf {\bibinfo {volume} {124}},\ \bibinfo
  {pages} {131103} (\bibinfo {year} {2006})}\BibitemShut {NoStop}%
\bibitem [{\citenamefont {O\u{g}uz}, \citenamefont {Messina},\ and\
  \citenamefont {L\"owen}(2011)}]{OML11}%
  \BibitemOpen
  \bibfield  {author} {\bibinfo {author} {\bibfnamefont {E.~C.}\ \bibnamefont
  {O\u{g}uz}}, \bibinfo {author} {\bibfnamefont {R.}~\bibnamefont {Messina}},\
  and\ \bibinfo {author} {\bibfnamefont {H.}~\bibnamefont {L\"owen}},\
  }\bibfield  {title} {\enquote {\bibinfo {title} {Helicity in cylindrically
  confined {Y}ukawa systems},}\ }\href
  {https://doi.org/10.1209/0295-5075/94/28005} {\bibfield  {journal} {\bibinfo
  {journal} {Europhys. Lett.}\ }\textbf {\bibinfo {volume} {94}},\ \bibinfo
  {pages} {28005} (\bibinfo {year} {2011})}\BibitemShut {NoStop}%
\bibitem [{\citenamefont {Montero}\ and\ \citenamefont
  {Santos}(2023{\natexlab{a}})}]{MS23b}%
  \BibitemOpen
  \bibfield  {author} {\bibinfo {author} {\bibfnamefont {A.~M.}\ \bibnamefont
  {Montero}}\ and\ \bibinfo {author} {\bibfnamefont {A.}~\bibnamefont
  {Santos}},\ }\bibfield  {title} {\enquote {\bibinfo {title} {Structural
  properties of hard-disk fluids under single-file confinement},}\ }\href
  {https://doi.org/%10.1063/5.0156228} {\bibfield  {journal} {\bibinfo
  {journal} {J. Chem. Phys.}\ }\textbf {\bibinfo {volume} {159}},\ \bibinfo
  {pages} {034503} (\bibinfo {year} {2023}{\natexlab{a}})}\BibitemShut
  {NoStop}%
\bibitem [{\citenamefont {Montero}\ and\ \citenamefont
  {Santos}(2024{\natexlab{a}})}]{MS24b}%
  \BibitemOpen
  \bibfield  {author} {\bibinfo {author} {\bibfnamefont {A.~M.}\ \bibnamefont
  {Montero}}\ and\ \bibinfo {author} {\bibfnamefont {A.}~\bibnamefont
  {Santos}},\ }\bibfield  {title} {\enquote {\bibinfo {title} {Exploring
  anisotropic pressure and spatial correlations in strongly confined hard-disk
  fluids. {E}xact results},}\ }\href
  {https://doi.org/10.1103/PhysRevE.110.L022601} {\bibfield  {journal}
  {\bibinfo  {journal} {Phys. Rev. E}\ }\textbf {\bibinfo {volume} {110}},\
  \bibinfo {pages} {L022601} (\bibinfo {year}
  {2024}{\natexlab{a}})}\BibitemShut {NoStop}%
\bibitem [{\citenamefont {Varga}(2025)}]{V25}%
  \BibitemOpen
  \bibfield  {author} {\bibinfo {author} {\bibfnamefont {S.}~\bibnamefont
  {Varga}},\ }\bibfield  {title} {\enquote {\bibinfo {title} {Positional
  ordering and close packing of hard spheres in nanochannels},}\ }\href
  {https://doi.org/10.1103/7spz-tppl} {\bibfield  {journal} {\bibinfo
  {journal} {Phys. Rev. E}\ }\textbf {\bibinfo {volume} {111}},\ \bibinfo
  {pages} {065418} (\bibinfo {year} {2025})}\BibitemShut {NoStop}%
\bibitem [{\citenamefont {Montero}\ and\ \citenamefont
  {Santos}(2024{\natexlab{b}})}]{MS24}%
  \BibitemOpen
  \bibfield  {author} {\bibinfo {author} {\bibfnamefont {A.~M.}\ \bibnamefont
  {Montero}}\ and\ \bibinfo {author} {\bibfnamefont {A.}~\bibnamefont
  {Santos}},\ }\bibfield  {title} {\enquote {\bibinfo {title} {Exact
  equilibrium properties of square-well and square-shoulder disks in
  single-file confinement},}\ }\href
  {https://doi.org/10.1103/PhysRevE.110.024601} {\bibfield  {journal} {\bibinfo
   {journal} {Phys. Rev. E}\ }\textbf {\bibinfo {volume} {110}},\ \bibinfo
  {pages} {024601} (\bibinfo {year} {2024}{\natexlab{b}})}\BibitemShut
  {NoStop}%
\bibitem [{\citenamefont {Montero}\ and\ \citenamefont
  {Santos}(2023{\natexlab{b}})}]{MS23}%
  \BibitemOpen
  \bibfield  {author} {\bibinfo {author} {\bibfnamefont {A.~M.}\ \bibnamefont
  {Montero}}\ and\ \bibinfo {author} {\bibfnamefont {A.}~\bibnamefont
  {Santos}},\ }\bibfield  {title} {\enquote {\bibinfo {title} {Equation of
  state of hard-disk fluids under single-file confinement},}\ }\href
  {https://doi.org/10.1063/5.0139116} {\bibfield  {journal} {\bibinfo
  {journal} {J. Chem. Phys.}\ }\textbf {\bibinfo {volume} {158}},\ \bibinfo
  {pages} {154501} (\bibinfo {year} {2023}{\natexlab{b}})}\BibitemShut
  {NoStop}%
\bibitem [{\citenamefont {Santos}(2016)}]{S16}%
  \BibitemOpen
  \bibfield  {author} {\bibinfo {author} {\bibfnamefont {A.}~\bibnamefont
  {Santos}},\ }\href@noop {} {\emph {\bibinfo {title} {A Concise Course on the
  Theory of Classical Liquids. Basics and Selected Topics}}},\ \bibinfo
  {series} {Lecture Notes in Physics}, Vol.\ \bibinfo {volume} {923}\ (\bibinfo
   {publisher} {Springer},\ \bibinfo {address} {New York},\ \bibinfo {year}
  {2016})\BibitemShut {NoStop}%
\bibitem [{\citenamefont {Maestre}\ \emph {et~al.}(2011)\citenamefont
  {Maestre}, \citenamefont {Santos}, \citenamefont {Robles},\ and\
  \citenamefont {{L\'opez de Haro}}}]{MSRH11}%
  \BibitemOpen
  \bibfield  {author} {\bibinfo {author} {\bibfnamefont {M.~A.~G.}\
  \bibnamefont {Maestre}}, \bibinfo {author} {\bibfnamefont {A.}~\bibnamefont
  {Santos}}, \bibinfo {author} {\bibfnamefont {M.}~\bibnamefont {Robles}},\
  and\ \bibinfo {author} {\bibfnamefont {M.}~\bibnamefont {{L\'opez de
  Haro}}},\ }\bibfield  {title} {\enquote {\bibinfo {title} {On the relation
  between virial coefficients and the close-packing of hard disks and hard
  spheres},}\ }\href {https://doi.org/10.1063/1.3558779} {\bibfield  {journal}
  {\bibinfo  {journal} {J. Chem. Phys.}\ }\textbf {\bibinfo {volume} {134}},\
  \bibinfo {pages} {{084}{502}} (\bibinfo {year} {2011})}\BibitemShut {NoStop}%
\bibitem [{\citenamefont {Montero}(2025)}]{SingleFileHardSpheres}%
  \BibitemOpen
  \bibfield  {author} {\bibinfo {author} {\bibfnamefont {A.~M.}\ \bibnamefont
  {Montero}},\ }\href@noop {} {\enquote {\bibinfo {title}
  {{SingleFileHardSpheres}},}\ }\bibinfo {howpublished}
  {\url{https://github.com/amonterouex/SingleFileHardSpheres}} (\bibinfo {year}
  {2025})\BibitemShut {NoStop}%
\bibitem [{\citenamefont {F\"uredi}(1991)}]{F91}%
  \BibitemOpen
  \bibfield  {author} {\bibinfo {author} {\bibfnamefont {Z.}~\bibnamefont
  {F\"uredi}},\ }\bibfield  {title} {\enquote {\bibinfo {title} {The densest
  packing of equal circles into a parallel strip},}\ }\href
  {https://doi.org/10.1007/BF02574677} {\bibfield  {journal} {\bibinfo
  {journal} {Discrete Comput. Geom}\ }\textbf {\bibinfo {volume} {6}},\
  \bibinfo {pages} {95--106} (\bibinfo {year} {1991})}\BibitemShut {NoStop}%
\end{thebibliography}

%

\end{document}